\documentclass[11pt]{article}
\usepackage{amsmath,amssymb,amsfonts,amsthm}
\usepackage[usenames,dvipsnames]{xcolor}
\usepackage{bm,xspace}
\usepackage{aliascnt}
\usepackage{fullpage}
\usepackage{liyang}
\usepackage{framed}
\usepackage{enumitem}
\usepackage{array}
\usepackage{multirow}
\usepackage{afterpage}
\usepackage{mathrsfs}  
\usepackage{caption}
\usepackage[normalem]{ulem}
\usepackage[backref, colorlinks,citecolor=blue,linkcolor=magenta,bookmarks=true]{hyperref}  
\usepackage[capitalise]{cleveref}
\crefname{claim}{Claim}{Claims}
\crefname{fact}{Fact}{Facts}
\crefname{coro}{Corollary}{Corollaries}

\theoremstyle{plain}

\newcommand{\dtv}{d_\mathrm{TV}}
\newcommand{\eps}{\varepsilon}
\newcommand{\okappa}{\overline{\kappa}}
\newcommand{\bmsg}{\boldsymbol{m}}
\newcommand{\sign}{\mathrm{sign}}
\newcommand{\R}{\mathbb R}

\newcommand{\N}{\mathbb N}

\newcommand{\ptf}[2]{\mathsf{PTF}^{#1}_{#2}}
\newcommand{\ltf}{\mathsf{LTF}}

\newcommand{\setOfSuchThat}[2]{ \left\{\; #1 \;\colon\; #2\; \right\} } 			
\newcommand{\norm}[1]{\lVert#1{\rVert}}
\newcommand{\normone}[1]{{\norm{#1}}_1}
\newcommand{\normtwo}[1]{{\norm{#1}}_2}
\newcommand{\norminf}[1]{{\norm{#1}}_\infty}
\newcommand{\abs}[1]{\left\lvert #1 \right\rvert}

\newcommand{\dotprod}[2]{ \left\langle #1,\xspace #2 \right\rangle } 

\newcommand{\Algo}{\ensuremath{\mathcal{A}}\xspace}
 
\newcommand{\class}{\ensuremath{\mathcal{C}}\xspace} 
\newcommand{\eqdef}{:=}

\newcommand{\proba}{\Pr}
\newcommand{\probaOf}[1]{\proba\!\left[\, #1\, \right]}

\newcommand{\probaDistrOf}[2]{\proba_{#1}\left[\, #2\, \right]}
\newcommand{\STAT}{{\sf STAT}\xspace}
\newcommand{\uniform}{\ensuremath{\mathcal{U}}}

\newcommand{\expect}[1]{\mathbb{E}\!\left[#1\right]}

\newcommand{\shortexpect}{\mathbb{E}}

\newenvironment{proofof}[1]{\begin{proof}[Proof of {#1}]}{\end{proof}}

\newenvironment{proofsketchof}[1]{\begin{proof}[Proof Sketch of {#1}]}{\end{proof}}

\newtheorem{assumption}{Assumption}

\newcommand{\littleO}[1]{{o\!\left( #1 \right)}}
\newcommand{\bigO}[1]{{O\left( #1 \right)}}

\newcommand{\bigTheta}[1]{{\Theta\left( #1 \right)}}
\newcommand{\bigOmega}[1]{{\Omega\left( #1 \right)}}

\providecommand{\poly}{\operatorname*{poly}}

\newcommand{\lp}[1][1]{\ell_{#1}}

\begin{document}

\title{Learning from satisfying assignments under continuous distributions}

\author{
  Cl\'{e}ment L. Canonne\footnote{Supported by a Motwani fellowship.  Some of this work was carried out while visiting Northwestern University.}\\
  Stanford University \and
  Anindya De\footnote{{Supported by NSF CCF-1926872 (transferred from CCF-1814706). Most of the work done while the author was at Northwestern University.}}\\
  University of Pennsylvania \and
  Rocco A. Servedio\footnote{Supported in part by NSF awards CCF-1814873 and CCF-1563155.}\\
  Columbia University
}

\maketitle

\begin{abstract}
What kinds of functions are learnable from their satisfying assignments? Motivated by this simple question,
we extend the framework of~\cite{DDS:15}, which studied the learnability of probability distributions over $\zo^n$ defined by the set of satisfying assignments to ``low-complexity'' Boolean functions, to Boolean-valued functions defined over \emph{continuous} domains. In our learning scenario there is a known ``background distribution'' $\calD$ over $\R^n$ (such as a known normal distribution or a known log-concave distribution) and the learner is given i.i.d. samples drawn from a target distribution $\calD_f$, where $\calD_f$ is $\calD$ restricted to the satisfying assignments of an \emph{unknown} low-complexity Boolean-valued function $f$.  The problem is to learn an approximation $\calD'$ of the target distribution $\calD_f$ which has small error as measured in total variation distance.

We give a range of efficient algorithms and hardness results for this problem, focusing on the case when $f$ is a low-degree \emph{polynomial threshold function} (PTF).  When the background distribution $\calD$ is log-concave, we show that this learning problem is efficiently solvable for degree-1 PTFs (i.e.,~linear threshold functions) but not for degree-2 PTFs.  In contrast, when $\calD$ is a normal distribution, we show that this learning problem is efficiently solvable for degree-2 PTFs but not for degree-4 PTFs.  Our hardness results rely on standard assumptions about secure signature schemes. 
\end{abstract}

\newpage

\setcounter{page}{1}

\section{Introduction}

Over the past decade or so there has been an explosion of research effort aimed at establishing rigorous results about the learnability of various classes of probability distributions.  A natural goal in this line of work is to understand how the complexity of learning an unknown high-dimensional distribution scales with the complexity of the distribution being learned.  This immediately leads to a question of how to measure the computational complexity of a high-dimensional distribution; needless to say, a range of different approaches are possible here.  In early work, Kearns \emph{et al.}~\cite{KMR+:94} proposed a model in which the complexity of a distribution $\calD$ over $\zo^n$ is measured in terms of the circuit complexity of an $m$-input, $n$-output Boolean circuit $C\colon \zo^m \to \zo^n$ which is such that $\calD$ is the output distribution when a uniform random string from $\zo^m$ is given as input to $C$ (i.e., $\calD = C(\calU)$).  This is a natural approach, but as Kearns \emph{et al.} showed, even depth-1 circuits of bounded fan-in OR gates lead to distributions $\calD$ which are hard to learn in this model.  Moreover, it is not clear how to extend the~\cite{KMR+:94} framework to distributions over the \emph{continuous} domain $\R^n$ as opposed to the Boolean cube $\zo^n.$

More recently \cite{DDS:15} took a different approach by considering ``low-complexity'' distributions $\calU_f$ over $\zo^n$ which are uniform over the set of satisfying assignments of a ``low-complexity'' Boolean function $f\colon \zo^n \to \{-1,1\}.$  They gave a range of algorithms and hardness results for learning distributions $\calU_f$ of this sort, where the unknown low-complexity  Boolean function is assumed to be a linear threshold function, a degree-2 polynomial threshold function, a DNF formula, a monotone width-2 CNF, and so on.  The main positive results of~\cite{DDS:15} showed that if $f$ is a linear threshold function or a DNF formula, then the distribution $\calU_f$ is learnable to total variation distance $\eps$ by an efficient algorithm (in time $\poly(n,1/\eps)$ for $f$ a linear threshold function, and in time $n^{O(\log(n/\eps))}$ for $f$ a $\poly(n)$-term DNF formula). \cite{DDS:15} also established a range of hardness results showing that these positive results are close to the limit of what can be achieved by efficient algorithms in their model.  They showed that under known constructions of  secure cryptographic signature schemes, degree-2 polynomial threshold functions over $\zo^n$ already give rise to distributions which cannot be learned in sub-exponential time.  (The same is true for distributions arising from satisfying assignments of monotone width-2 CNF formulas.)

\paragraph{This work:  Learning continuous distributions from satisfying assignments}  One attractive feature of the~\cite{DDS:15} framework is that it can be straightforwardly extended to learning low-complexity distributions over \emph{continuous} domains.  Given a known ``background distribution'' $\calD$ (analogous to the uniform distribution over $\zo^n$) such as a normal distribution or a log-concave distribution over $\R^n$, one can naturally consider the problem of learning the distribution $\calD_f$ induced by restricting $\calD$ to the satisfying assignments of an unknown low-complexity Boolean-valued function $f\colon \R^n \to \zo$.  It is particularly natural in this setting to take $f$ to be a \emph{low-degree polynomial threshold function (PTF)}, as such functions (unlike, say, DNF or CNF formulas) can very naturally be viewed as functions over the entire domain $\R^n$.  (Recall that a degree-$d$ PTF over $\R^n$ is a function $f\colon \R^n \to \bits$ defined by $f(x) = \sign(p(x))$ where $p(x_1,\dots,x_n)$ is a degree-$d$ real polynomial.)  

With this motivation, and the additional motivating hope of establishing efficient learnability of distributions of satisfying assignments over $\R^n$ whose discrete analogues over $\zo^n$ do not have efficient learning algorithms, in this paper we consider the aforementioned extension of the~\cite{DDS:15} framework to learning \emph{continuous} distributions of satisfying assignments.  We define our learning framework, and describe our results in this framework, below.  

\subsection{The learning framework}   \label{sec:learning-framework}

We now introduce our learning model in more detail but still at an informal level; see~\cref{sec:preliminaries} for a detailed description of the model.  Similar to~\cite{DDS:15}, a learning problem in our framework is defined by a class $\calC$ of Boolean-valued functions from $\R^n$ to $\bits$.  In our setting an instance of the learning problem is defined by a choice of a function $f \in \calC$, which is \emph{unknown} to the learning algorithm, and a \emph{background distribution} $\calD$, which is provided to the learning algorithm.  (We explain in detail just what this means in \Cref{sec:preliminaries}; here we note that  the distribution $\calD$ is analogous to the uniform distribution over $\zo^n$, which is the background distribution in~\cite{DDS:15} and is certainly ``known'' to any learning algorithm.)  The learning algorithm has access to i.i.d.~samples drawn from $\calD_f$, which is the distribution $\calD$ restricted to the set of satisfying assignments $\setOfSuchThat{x \in \R^n }{ f(x)=1 }$ of $f$.  The learner's task is to output a hypothesis distribution $\calD'$ (we will amplify on precisely what this means in the next section) which is such that the total variation distance $\dtv(\calD_f,\calD')$ is at most $\eps$; a successful learning algorithm is one which does this, with high probability over its random samples and any internal randomness, for any $f \in \calC$.  We say that such an algorithm is a \emph{distribution learning algorithm for class $\calC$ with respect to background distribution $\calD$.}

\subsection{Our results}

The above framework is quite general; the results of \cite{DDS:15} fall into this framework by taking the background distribution ${\cal D}$ to be the uniform distribution over $\zo^n$.  It is not difficult to see that if the distribution ${\cal D}$ is totally unrestricted, then learning is intractable in this framework even for simple classes of Boolean functions such as linear threshold functions(see~\Cref{ap:intractable}).  Thus, in this paper we restrict our attention to two different types of background distributions $\calD$:  (i)~log-concave distributions over $\R^n$, and (ii)~normal distributions over $\R^n$.  (We recall that a distribution $\calD$ over $\R^n$ is log-concave if its density function $p$ satisfies $p = \exp(\phi)$ where $\phi\colon \R \to [-\infty,\infty)$ is a concave function.)  We further restrict our attention to the function class $\calC = \ptf{n}{d}$, the class of polynomial threshold functions over $\R^n$ of degree at most $d$, where $d$ is some fixed small constant.
\medskip

\noindent\textbf{Log-concave distributions.}  Our first positive result is an efficient distribution learning algorithm for linear threshold functions (i.e., PTFs of degree 1) with respect to any log-concave background distribution:

\begin{theorem} [Informal statement; see~\cref{thm:pos-deg1-detailed} in~\cref{sec:preliminaries} for detailed statement] \label{thm:pos-deg1}
There is an algorithm $A$, running in time $\poly(n,1/\eps)$, that is a distribution learning algorithm for the class $\ltf$ of $n$-variable linear threshold functions with respect to any log-concave background distribution $\calD$ over $\R^n$.
\end{theorem}

\noindent We complement this positive result with a negative result for degree-2 PTFs, showing that the above is essentially best possible for log-concave background distributions:
\begin{theorem} \label{thm:neg-deg2}
Under known constructions of secure signature schemes, there is no subexponential-time algorithm $A$ for learning the class $\ptf{n}{2}$ of $n$-variable degree-two polynomial threshold functions with respect to every log-concave background distribution $\calD$.
\end{theorem}

\noindent\textbf{Normal distributions.}  Turning next to normal distributions, our first positive result is an efficient distribution learning algorithm for degree-two polynomial threshold functions with respect to any normal distribution:

\begin{theorem} [Informal statement; see~\cref{thm:pos-deg2-detailed} in~\cref{sec:preliminaries} for detailed statement] \label{thm:pos-deg2}
There is an algorithm $A$, running in time $\poly(n,1/\eps)$, that is a distribution learning algorithm for the class $\ptf{n}{2}$ of $n$-variable degree-two polynomial threshold functions with respect to any normal background distribution $\calD$ over $\R^n$.
\end{theorem}

\noindent We complement this positive result with a negative result for degree-4 PTFs:

\begin{theorem} \label{thm:neg-deg4}
Under known constructions of secure signature schemes, there is no subexponential-time algorithm $A$ for learning the class $\ptf{n}{4}$ of $n$-variable degree-four polynomial threshold functions with respect to the standard normal distribution ${\cal N}(0,1)^n$.
\end{theorem}

\noindent \textbf{Discussion.}  \cref{thm:pos-deg1,thm:neg-deg2} align closely with the positive and negative results established in~\cite{DDS:15} in the discrete setting where the background distribution $\calD$ is simply the uniform distribution $\calU$ over $\zo^n$.  Recall that in~\cite{DDS:15} it is shown that there is an efficient distribution learning algorithm for the class of linear threshold functions with respect to the background distribution $\calU$, but there is no such efficient distribution learning algorithm for degree-2 PTFs with respect to the background distribution $\calU$.  In contrast,~\cref{thm:pos-deg2} shows that normal distributions are ``easier'' background distributions than uniform or arbitrary log-concave distributions, as satisfying assignments of degree-2 PTFs \emph{are} efficiently learnable for normal background distributions.  But there is a limit to the power afforded by normal background distributions, as~\cref{thm:neg-deg4} shows that already degree-4 PTFs are hard to learn under normal background distributions.

\subsection{Related work} 

A recent line of work which focuses on learning high-dimensional normal distributions from truncated samples~\cite{DGTZ18,KTZ19} has a somewhat similar flavor to ours: in this setting, the underlying distribution is an unknown multivariate normal $\calD=\calN(\mu,\Sigma)$, and observed samples come from $\calD_f$, where $f$ is the indicator function of an arbitrary truncation set $S\subseteq \R^n$. The goal is to learn the parameters $\mu$ and $\Sigma$ of the \emph{true} distribution (or, equivalently, to learn $\calD$).
Daskalakis, Gouleakis, Tzamos, and Zampetakis~\cite{DGTZ18}  gave an efficient algorithm for this task when the truncation set $S$ is known. An incomparable result of Kontonis, Tzamos, and Zampetakis~\cite{KTZ19} gives an efficient algorithm
to recover $\mu$ and $\Sigma$ 
even when $S$ is unknown; however, \cite{KTZ19} require $S$ to have bounded surface area (see~\cite{KOS:08} for background on this notion) and, more importantly, the covariance matrix $\Sigma$ is required to be diagonal.

In contrast, in our setting the parameters $\mu$ and $\Sigma$ are known to the algorithm; the function $f$ is assumed to be an unknown LTF or degree-$2$ PTF (for our positive results); and the goal is to learn the truncated distribution ${\cal D}_f$. One important difference between our results and those of \cite{DGTZ18,KTZ19} is that our algorithm is efficient even when the truncation set $S$ (i.e.,~the set of satisfying assignments of the unknown LTF or PTF) has inverse exponential density in the Gaussian space; indeed, achieving this is the most challenging aspect of our result. In contrast, in the setting of \cite{DGTZ18, KTZ19}, even achieving a sample-efficient algorithm requires that the truncation set $S$ has at least an inverse $\poly(n)$ density. Further, for a computationally efficient algorithm, \cite{KTZ19} requires that the truncation set has a constant density in the Gaussian space. 
 
Finally, another work which is  related to the current paper is that of \cite{AGR13}, which gives an efficient algorithm for learning an $n$-dimensional simplex when the algorithm is given uniformly random samples from (the uniform distribution on) the simplex. Recall that a simplex in $\mathbb{R}^n$ is the same as the intersection of exactly $n+1$ halfspaces along with the requirement that the intersection is a bounded set. Their result and tools seem quite different from ours (as their techniques crucially rely on a reduction to ICA).

\subsection{Our upper bound techniques}

At the highest level our approach follows the general method for learning from satisfying assignments
that was given in~\cite{DDS:15}, though as described later there are many technical issues and obstacles
that are specific to this work. \cite{DDS:15} developed a general framework in which, very roughly, if
one is given a number of different types of algorithms for a class $\calC$ of functions from $\zo^n$ to
$\zo$, then one obtains a distribution learning algorithm for $\calC$ with respect to the background
distribution $\calU$, the uniform distribution over the Boolean hypercube $\zo^n$. The required algorithms are specifically (i)~an approximate uniform generation algorithm for $\calC$, (ii)~an approximate
counting algorithm for $\calC$, (iii)~a Statistical Query learning algorithm for $\calC$, and (iv) a ``densifier''
for $\calC$, explained below.  (See Theorem 3.1
of~\cite{DDS:15} for a precise statement of the original framework.) We note that approximate uniform
generation, approximate counting, and statistical query learning are of course well-studied and
standard notions, but the notion of a ``densifier'' was new to~\cite{DDS:15} and a significant amount
of the technical work in that paper came in developing densifiers for  linear threshold functions and DNF formulas.

As our starting point, we observe that the algorithmic framework of~\cite{DDS:15} can be straightforwardly extended to \emph{non-uniform} background distributions $\calD$ by making some natural modifications. In particular,
\begin{enumerate}
  \item  In place of an approximate uniform generation algorithm (which outputs approximately uniform satisfying assignments of a function $f \in \calC$) one needs the natural analogue for
distribution $\calD$, namely a procedure for \emph{approximately sampling} from $\calD_f$.
  \item  In place of an approximate counting algorithm for $\calC$ (which outputs an approximation of $\abs{f^{-1}(1)}/2^n$, where $f\colon \zo^n \to \zo$ belongs to $\calC$), one needs a procedure for \emph{approximate integration} of $\calD$ over the domain $f^{-1}(1)$.
  \item  We also need a \emph{densifier with respect to $\calD$}, which is a natural extension of the densifier notion from~\cite{DDS:15}.
  Roughly speaking, a densifier is an algorithm which is given random positive examples of $f$ (drawn according to ${\cal D}$ restricted to $f^{-1}(1)$) and prunes the entire domain $\R^n$ down to a set $S$ which (essentially) contains all of $f^{-1}(1)$ and which is such that $f^{-1}(1)$ is ``dense'' in $S$ under distribution ${\cal D}$.
  
\end{enumerate}

(The fourth ingredient, a Statistical Query algorithm, is unchanged from~\cite{DDS:15} to the present
work, since even the earlier work needed Statistical Query learning with respect to various non-uniform distributions.) In~\cref{sec:upperbounds} we give a precise statement of the extension of~\cite{DDS:15}'s Theorem 3.1,~\cref{thm:generalmethod}, which we will use to obtain our results. In the rest of this subsection we
give an overview of how each of the above three ingredients~--~approximate sampling, approximate
integration, and densification, all with respect to $\calD$~--~are achieved for each of our two positive
results (learning linear threshold functions with respect to arbitrary log-concave distributions, and
learning degree-2 PTFs with respect to the normal distribution). We begin with the simpler scenario
of linear threshold functions.

\medskip
\noindent {\bf Learning LTFs with respect to log-concave distributions.} The key insight here is that if
$\calD$ is a log-concave distribution and $f$ is any LTF, then since $f^{-1}(1)$ is a convex set in $\R^n$, the
distribution $\calD_f$ is again a log-concave distribution. Our approach leverages this simple observation
using known results, due to~\cite{LV06}, for approximate sampling and approximate
integration of log-concave distributions. (We note that significant work is required to align these known results with our setting; see~\Cref{sec:logconcave-sample-integrate}, where~\Cref{lem:sampling-our-assump} formally establishes the approximate sampling and approximate integration results that we require.)

For the densifier, at a high level our approach is similar
to the construction of a densifier for linear threshold functions in~\cite{DDS:15} where the background
distribution is uniform over $\zo^n$. Like~\cite{DDS:15}, we show how an efficient \emph{online mistake-bound}
algorithm for linear threshold functions can be leveraged to obtain the desired densifier, now with
respect to the log-concave background distribution $\calD$; however, subtleties arise in our current setting which were not present in~\cite{DDS:15} since we are now working in a continuous rather than discrete setting. In particular, online algorithms for learning LTFs with a finite mistake bound  do not exist over continuous domains. To circumvent this issue, we use anticoncentration properties of log-concave distributions and the fact that the boundary of $f^{-1}(1)$ is a very ``well-structured'' set (it is simply a hyperplane in $\R^n$)  to give an analysis in~\Cref{sec:upperbounds:1ptf:logconcave} showing that, after suitable discretization of the domain, one can apply the discrete-setting mistake-bound algorithms without compromising the required guarantees.  Intuitively, this is because a suitably fine discretization only ``flips the label'' of a tiny fraction of points.

\medskip
\noindent {\bf Learning degree-2 PTFs with respect to normal distributions.} Conceptually, the densification step
for degree-2 PTFs and normal distributions can be handled in roughly the same way as the densification step for the LTF scenario described above.  (There are some differences in the technical arguments which show that  online mistake-bound learning algorithms can be successfully used in the discretized version of the degree-2 PTF setting.  In this setting the boundary of $f^{-1}(1)$ is not simply a hyperplane, so now we must rely  on results of Carbery and Wright~\cite{CW:01} which establish anticoncentration of quadratic polynomials in Gaussian random variables; details are in~\Cref{sec:upperbounds:2ptf:gaussian:densifier}.)  However, new challenges arise in
obtaining the required approximate sampling and approximate integration algorithms. The central
obstacle is that for $f$ a degree-2 PTF over $\R^n$, the set $f^{-1}(1)$ can be non-convex; this means that
the resulting restricted normal distribution $\calN_f$ need not be log-concave, so the known tools for
working with log-concave distributions, such as~\cite{LV06}, are no longer at our disposal.
We circumvent this obstacle through a careful approach which heavily exploits the degree-2 polynomial
structure of $f$.\footnote{This should not be a surprise, recalling our negative results for degree-4 polynomial threshold functions.}

We first discuss the approximate integration/counting problem. 
Via a change of basis, we convert a general
degree-2 polynomial to a ``decoupled'' degree-2 multivariate polynomial, which is a sum of $n$ 
univariate quadratic polynomials over distinct variables, of the form
\[
\sum_{i=1}^n \lambda_i (x_i - \mu_i)^2 + c.
\]
This change of basis does not change the underlying
geometry of the problem, but it enables us to reframe the counting/integration problem as the problem of estimating the probability that a random Gaussian $\bG = (\bG_1,\dots,\bG_n)$ satisfies a linear inequality
\[
\sum_{i=1}^n \lambda_i (\bG_i - \mu_i)^2 \geq C.
\]
Crucially, the above polynomial is a sum of \emph{independent} random variables $\lambda_i (\bG_i - \mu_i)^2$.
We solve this problem by adapting a FPTAS, due to Li and Shi \cite{LS:14}, for multiplicatively approximating the probability that a sum of independent random variables exceeds a threshold value.  The results of \cite{LS:14} are for discrete random variables, whereas our setting is continuous; in our adaptation we use anticoncentration results due to Carbery and Wright \cite{CW:01} in order to establish that the rounding we perform (so that the \cite{LS:14} results can be used) only incurs an acceptably low amount of error.  Details are given in~\Cref{sec:upperbounds:2ptf:gaussian:counting}.

It remains only to give an efficient routine for approximate sampling from the normal distribution subject to a degree-2 PTF constraint.  This is done in~\Cref{sec:upperbounds:2ptf:gaussian:sampling}, where we extend the usual reduction from approximate counting to approximate sampling and leverage our approximate counting result described in the previous paragraph; an extension is required because the usual reduction is for counting and sampling problems over the domain $\zo^n$, whereas now the relevant domain is $\R^n$.  We show that it suffices to discretize $\R^n$ to an exponentially fine grid, and augment the usual reduction with binary search in each coordinate (which can be carried out efficiently even over an exponential-size domain).

We note that, prior to this work, no (provably) efficient algorithm was known for sampling from a high-dimensional normal subject to quadratic constraints, a task used as a subroutine in many applied settings (see, e.g.,~\cite{EllisM07,pakman2014exact}, and references therein).

\subsection{Our lower bound techniques}

Our computational hardness results are obtained via an extension of the approach that was given in~\cite{DDS:15}. This approach is based on a connection between secure cryptographic signature schemes
with certain properties and learning from satisfying assignments. The crux of the idea is very simple
when the background distribution is uniform over $\zo^n$ (as in~\cite{DDS:15}): if we view $\calU_f$ as the
uniform distribution over signed messages, an algorithm which can construct a sampler for $\calU_f$ given
access to independent draws from it corresponds to an algorithm which, given a batch of signed
messages, can generate a new signed message, and this violates the security of a signature scheme.
Via reductions, this intuition is leveraged to show that assuming the existence of suitably secure
signature schemes, no efficient algorithm can learn various types of functions given access only to
random satisfying assignments.

In the current paper we extend this connection to our setting of more general background
distributions other than uniform on $\zo^n$. The heart of our lower bound proof for degree-2
PTFs under log-concave distributions is a reduction from \textsf{Subset-Sum}. By designing a suitable
degree-2 PTF based on a subset sum instance, we are able to establish hardness even when the
background distribution is a very simple and well-structured log-concave distribution, namely the
uniform distribution on the solid cube $[0, 1]^n$; this gives~\cref{thm:neg-deg2}.\footnote{Some care is required to ensure that our lower bound construction complies with a natural ``reasonableness''
requirement we put on learning problems in our framework, which is that the set of all satisfying assignments $f^{-1}(1)$ has at least an inverse exponential amount of probability mass under the background distribution~--~see the discussion in~\Cref{sec:preliminaries}~--~but this is essentially a technical issue.}

Since the background log-concave distribution for our degree-2 PTF hardness result is uniform on $[0,1]^n$, which
is very ``tame,'' it is natural to wonder whether this hardness result can be extended to the normal
distribution $\calN(0,1)^n$ as the background distribution. However, this natural intuition is directly
contradicted by our main algorithmic result,~\Cref{thm:pos-deg2}, which says that degree-2 PTFs are \emph{easy} to
learn under the normal background distribution! Indeed, it turns out that the bounded support of
$[0,1]^n$ plays an essential role in letting the reduction go through. We show that by working with a
suitable degree-\emph{four} polynomial, it is possible to essentially enforce the bounded support requirement
even when the background distribution is $\calN(0,1)^n$, and we thereby obtain computational hardness
for learning degree-four PTFs with respect to the normal background distribution. The learnability
of degree-three PTFs with respect to the normal background distribution is an interesting open
problem for future work.

\section{Preliminaries} \label{sec:preliminaries}

\paragraph{Detailed description of our model.}  We now explain in detail the sense in which, as stated in~\Cref{sec:learning-framework},   the background distribution $\calD$ is ``provided to''  the learning algorithm.  

For the case where $\calD$ is a normal distribution, this simply means that the learning algorithm is given, as part of its input, the mean $\mu \in \R^n$ and covariance matrix $\Sigma \in \R^{n \times n}$ defining the normal distribution.  In fact, since the class of low-degree polynomial threshold functions is closed under affine transformations of the input space $\R^n$ (i.e.,~if $f(x)$ is a PTF of degree $d$, then $f(Ax + b)$ is also a PTF of degree $d$ for every $n \times n$ matrix $A$ and every vector $b \in \R^n$), for our normal distribution results we may simply assume that the background distribution is the standard $\calN(0,1)^n$ distribution.  (If the background normal distribution is not full dimensional, then an affine transformation converts it to $\calN(0,1)^k$ for some $k < n$, and our algorithms can simply be carried out over $\R^k$ rather than $\R^n.$)  Thus we subsequently assume, without loss of generality for our normal distribution results, that the given normal background distribution is simply $\calN(0,1)^n$.

For the case when $\calD$ is a log-concave distribution, the learning algorithm is provided with an evaluation oracle for a function $\tilde{\mu}$ which is proportional to the probability density function $\mu\colon \R^n\to[0,1]$ of $\calD$. We will assume that $\tilde{\mu}$ satisfies three conditions, where $C>1$ is some sufficiently large absolute constant: 

\begin{itemize}
\item[\textbf{C1:}] all but at most $2^{-Cn}$ of the probability mass is concentrated in the $L_2$ ball of radius $n^C$; 

\item[\textbf{C2:}] $\norminf{\mu}\leq 2^n$ and $\norminf{\tilde{\mu}} \leq 2^{n^C}$;

\item[\textbf{C3:}]  Let $\tilde{\mu}_T(x)$ be  obtained by truncating $\tilde{\mu}(x)$ at $n^C$ bits after the binary point. Then
\[
\int \frac{1}{2} \tilde{\mu}(x) \le \int \tilde{\mu}_T(x) \le\int \tilde{\mu}(x) dx\,.
\]
\end{itemize}

The first two conditions restrict to log-concave distributions that are at least mildly concentrated, and not too ``peaked,'' while the third condition says that truncating $\tilde{\mu}$ up to polynomial precision provides a good approximation (in a certain weak average sense).  We note that these three conditions are quite benign and do not significantly restrict the scope of our results. In particular, they are satisfied by any uniform distribution over a convex set of reasonable size (as well as by the distributions used in our hardness results).
We refer to a log-concave distribution that satisfies the above conditions {\bf C1} through {\bf C3} as being \emph{well-behaved}.

Next we specify what is exactly meant for the learning algorithm to ``output a hypothesis distribution.'' What we require is that the learning algorithm output a description of a polynomial-time procedure which has query access to $\tilde{\mu}$ (equivalently, it outputs a polynomial-size circuit with oracle gates for $\tilde{\mu}$) which, on input uniformly random bits, outputs a draw from a distribution $\calD'$ that is $\eps$-close to $\calD_f$.

\medskip
\noindent {\bf Detailed theorem statements.} 
Before giving detailed theorem statements, we briefly discuss some numerical issues that arise since we are considering distributions over the continuous space $\R^n$.  If the target function $f\colon \R^n \to \bits$ and background distribution $\calD$ over $\R^n$ are such that $f^{-1}(1)$ has extremely low mass under $\calD$ --- say, $\Pr_{\bx \gets \calD}[f(\bx)=1] = 2^{-n^{\omega(1)}}$ --- then a super-polynomial number of bits may be required even to differentiate any point in $f^{-1}(1)$ from points in $f^{-1}(-1).$  To avoid such extreme scenarios, we assume for our positive results that the target function $f$ has a  ``reasonable'' (at least inverse exponential) fraction of satisfying assignments under the background distribution.  Thus, a detailed statements of our positive results is as follows:

\begin{theorem} [Detailed statement of~\Cref{thm:pos-deg1}:  algorithm for LTFs under log-concave background distribution] \label{thm:pos-deg1-detailed}
There is an algorithm $A$ with the following property:  Let $\calD$ be a well-behaved log-concave distribution provided to $A$ as described above, and let $f$ be a unknown LTF over $\R^n$ such that $\Pr_{\bx \gets \calD}[f(\bx)=1] \geq 2^{-n}.$  Given access to i.i.d.~samples from $\calD_f = $ ($\calD$ restricted to $f^{-1}(1)$), for $\eps = 1/2^{o(n)}$ algorithm $A$ runs in $\poly(n,1/\eps)$ time and with high probability outputs a hypothesis distribution $\calD'$ such that $\dtv(\calD',\calD_f) \leq \eps.$
\end{theorem}

\begin{theorem} [Detailed statement of~\Cref{thm:pos-deg2}:  algorithm for degree-2 PTFs under normal background distribution] \label{thm:pos-deg2-detailed}
There is an algorithm $A$ with the following property:  Let $f$ be a unknown degree-two PTF over $\R^n$ such that $\Pr_{\bx \gets {\cal N}(0,1)^n}[f(\bx)=1] \geq 2^{-n}.$  Given access to i.i.d.~samples from $\calN_f =$ ($\calN(0,1)^n$ restricted to $f^{-1}(1)$), for $\eps = 1/2^{o(n)}$ algorithm $A$ runs in $\poly(n,1/\eps)$ time and with high probability outputs a hypothesis distribution $\calD'$ such that $\dtv(\calD',\calN_f) \leq \eps.$
\end{theorem}

\begin{remark} \label{remark:reasonable}
We remark that our hardness results as well as our algorithms will adhere to this ``reasonableness'' condition: in our constructions establishing computational hardness, the learning problems which we show to be hard are all ones in which the target function $f$ and background distribution $\calD$ are such that $\Pr_{\bx \leftarrow \calD}[f(\bx)=1]$ is at least $2^{-\poly(n)}.$
\end{remark}

\subsection{The framework of~\cite{DDS:15} and the ingredients it requires}
We now recall the algorithmic components of~\cite{DDS:15} which, suitably generalized to the continuous case, will prove instrumental to our upper bounds. In order to do so, we first must introduce some of the notions they require, starting with that of a \emph{densifier}:
\begin{definition}
Fix a function $\gamma(n,1/\eps,1/\delta)$ taking values in $(0,1]$ and two classes $\class,\class'$ of $n$-variate Boolean functions from $\R^n$ to $\bits$. An algorithm $\Algo_{\rm den}^{(\class, \class')}$ is said to be a \emph{$\gamma$-densifier for $\class$ using $\class'$ under $\calD$} if it has the following behavior: For every $\eps,\delta\in(0,1]$, every $1/2^n \leq \hat{p}\leq 1$, and every $f\in\class$, given as input $\eps,\delta,\hat{p}$ and a set of independent samples from $\calD_f$, the following holds. Let $p\eqdef \probaDistrOf{\bx \leftarrow\calD}{f(\bx)=1}$. If $p\leq \hat{p} < (1+\eps)p$, then with probability at least $1-\delta$ the algorithm outputs a function $g\in\class'$ such that:
(a) $\probaDistrOf{\bx\leftarrow\calD_f}{g(\bx)=1} \geq 1-\eps$, and 
(b) $\probaDistrOf{\bx\leftarrow\calD_g}{f(\bx)=1} \geq \gamma(n,1/\eps,1/\delta)$.
\end{definition}

The intuition behind this definition is that by (a) almost all of the satisfying assignments of $f$ are also satisfying assignments of $g$, and by (b) the satisfying assignments of $f$ are ``dense'' in $g^{-1}(1)$ (at least a $\gamma$ fraction under ${\cal D}$).  We will sometimes make explicit the parameters of the definition, and write that such an algorithm is an $(\eps,\gamma,\delta)$-densifier. 

The second notion we need is that of the \emph{statistical query} (SQ) learning model from~\cite{Kearns:98}, a restriction of the PAC learning model which only allows the learning algorithm to obtain, via a statistical query oracle $\STAT(f,\calD)$, estimates of the form $\shortexpect_{\bx\leftarrow\calD}[ \chi(\bx,f(\bx)) ]$ to within an additive $\tau$, where $\chi\colon\R^n\times\{-1,1\}\to\{-1,1\}$ and $\tau\in(0,1]$ constitute the query $(\chi,\tau)$, and $f\in\class$ is the unknown concept. As in the PAC model, the algorithm must output a hypothesis $g$ such that $\probaDistrOf{\bx\leftarrow\calD}{f(\bx)\neq g(\bx)} \leq \eps$ with probability at least $1-\delta$; such an algorithm, referred to as an \emph{$(\eps,\delta)$-SQ learning algorithm for $\class$}, is said to be efficient if it makes a number of queries polynomial in $n$, $1/\eps$, and $\log(1/\delta)$, all with accuracy $\tau=\poly(1/n,\eps)$. 

We will also need the standard definitions of \emph{approximate counting} and \emph{approximate generation}, slightly adapted to our setting in which the relevant background distribution is some other distribution than the uniform distribution over $\zo^n$.  Since it is more straightforward, we begin with the notion of approximate counting that we will require:
\begin{definition}
Let $\class$ be a class of $n$-variate Boolean functions mapping $\R^n$ to $\bits$. A randomized algorithm $\Algo_{\rm count}^{(\class)}$ is said to be an \emph{efficient approximate counting algorithm for $\class$ under $\calD$} if, for any $\eps,\delta\in(0,1]$ and any $f\in\class$, on input $\eps,\delta$ and $f\in\class$ and given oracle access to $\calD$, it runs in time $\poly(n,1/\eps,\log(1/\delta))$ and with probability at least $1-\delta$ outputs a value $\hat{p}$ such that
\[
    \frac{1}{1+\eps}\probaDistrOf{\bx\leftarrow\calD}{f(\bx)=1} \leq \hat{p} \leq (1+\eps)\probaDistrOf{\bx\leftarrow\calD}{f(\bx)=1}\,.
\]
\end{definition}

Taking ${\calD}$ to be the uniform distribution over $\zo^n$, this corresponds precisely to the usual notion of approximate counting which was used in \cite{DDS:15}.

There is an issue which arises in defining approximate generation algorithms in our setting which we now discuss (and explain how to handle).  We begin with the following definition:

\begin{definition} \label{def:generation}
Let $\class$ be a class of $n$-variate Boolean functions mapping $\R^n$ to $\bits$. A randomized algorithm $\Algo_{\rm gen}^{(\class)}$ is said to be an \emph{efficient strong approximate generation algorithm for $\class$ under $\calD$} if, for any $\eps\in(0,1]$ and $f\in\class$, there is a distribution $\hat{\calD}_{f,\eps}$ supported on $f^{-1}(1)$ with
\[
      \frac{1}{1+\eps}\calD_f(x) \leq \hat{\calD}_{f,\eps}(x) \leq (1+\eps)\calD_f(x)
\]
for every $x\in f^{-1}(1)$, such that for any $\delta\in(0,1]$, on input $\eps,\delta$, and $f\in\class$, and given oracle access to $\calD$, the algorithm runs in time $\poly(n,1/\eps,\log(1/\delta))$ and either outputs a point $x\in f^{-1}(1)$ exactly distributed according to $\hat{\calD}_{f,\eps}$, or outputs $\bot$. Moreover, the probability that it outputs $\bot$ is at most~$\delta$.
\end{definition} 

Taking the distribution ${\cal D}$ in \Cref{def:generation} to be the uniform distribution over $\bn$, we recover precisely the usual notion of approximate generation which was used in \cite{DDS:15}. While this compatibility with \cite{DDS:15} is an attractive feature, it is not difficult to see that the requirement that $\hat{\calD}_{f,\eps}$ put approximately the right amount of weight (in a multiplicative sense) on \emph{every} $x \in f^{-1}(1)$ is overly demanding.  If ${\cal D}_f$ puts only an extremely small amount of weight on a point $x$, it may be impossible for an efficient algorithm to very precisely match this extremely small amount of weight;  moreover, since the weight on $x$ is so tiny, as long as a purported approximate generation algorithm puts only a tiny amount of weight on $x$, it should not matter whether or not the tiny amount it puts is multiplicatively accurate.  We thus introduce the following relaxed notion of approximate generation, which we call ``weak approximate generation'':

\begin{definition} \label{def:weakgeneration}
Let $\class$ be a class of $n$-variate Boolean functions mapping $\R^n$ to $\bits$. A randomized algorithm $\Algo_{\rm gen}^{(\class)}$ is said to be an \emph{efficient weak approximate generation algorithm for $\class$ under $\calD$} if, for any $\tau \in(0,1]$ and any $f\in\class$, there is a distribution ${\calD}'_{f,\tau}$ with
\[
\dtv(\calD_f,{\calD}'_{f,\tau}) \leq \tau
\]
such that on input $\tau$, and $f\in\class$, and given oracle access to $\calD$, the algorithm runs in time $\poly(n,\log(1/\tau))$ and outputs a point $x\in f^{-1}(1)$ distributed according to ${\calD}'_{f,\tau}$. 
\end{definition}

\subsection{The algorithmic result we use} \label{sec:alg-result}
With the above notions in hand, we are ready to state the algorithmic result that we will rely on:
\begin{theorem}[based on {\cite[Theorem 3.1]{DDS:15}}] \label{thm:generalmethod}
  Let $\class,\class'$ be classes of $n$-variate Boolean functions mapping $\R^n$ to $\bits$. Suppose that 
  \begin{itemize}
    \item $\Algo_{\rm den}^{(\class, \class')}$ is an $(\eps,\gamma,\delta)$-densifier for $\class$ using $\class'$ under $\calD$ running in time $T_{\rm den}(n,1/\eps,1/\delta)$.
    \item $\Algo_{\rm count}^{(\class')}$ is an $(\eps,\delta)$-approximate counting algorithm for $\class'$ under $\calD$ running in time $T_{\rm count}(n,1/\eps,1/\delta)$.
    \item $\Algo_{\rm gen}^{(\class')}$ is an $(\eps,\delta)$-weak approximate generation algorithm for $\class'$ under $\calD$ running in time $T_{\rm gen}(n,1/\eps,1/\delta)$.\footnote{Actually, the theorem only requires that $\Algo_{\rm count}^{(\class')}$ and $\Algo_{\rm gen}^{(\class')}$ succeed as long as $f \in \mathcal{C'}$ satisfies $\Pr_{x \leftarrow \mathcal{D}} [f(x)=1] \ge 2^{-\poly(n)}$.}
    \item $\Algo_{\rm SQ}^{(\class)}$ is an $(\eps,\delta)$-SQ learning algorithm for $\class$ such that: $\Algo_{\rm SQ}^{(\class)}$ runs in time $t_1(n,1/\eps,1/\delta)$, $t_2(n)$ is the maximum time needed to evaluate any query provided to $\STAT$, and $\tau(n,1/\eps)$ is the minimum value of the tolerance parameter ever provided to $\STAT$ in the course of $\Algo_{\rm SQ}^{(\class)}$'s execution.
  \end{itemize}
  Then there exists a distribution learning algorithm $\Algo^{\class}$ for $\class$ under $\calD$. The running time of $\Algo^{\class}$ is polynomial in $T_{\rm den}(n,1/\eps,1/\delta)$, $T_{\rm count}(n,1/\eps,1/\delta)$, $T_{\rm gen}(n,1/\eps,1/\delta)$, $t_1(n,1/\eps,1/\delta)$, $t_2(n)$, and $1/\tau(n,1/\eps)$, provided that $T_{\rm den}(\cdot), T_{\rm count}(\cdot), T_{\rm gen}(\cdot), t_1(\cdot), t_2(\cdot), 1/\tau(\cdot)$ are polynomial in their parameters. 
\end{theorem}

The above statement is almost verbatim from~\cite{DDS:15}, with the difference that (i) we apply it to our setting of continuous distributions and Boolean functions over $\R^n$, and (ii) the third bullet uses the notion of a weak approximate generation algorithm rather than a strong one.~\Cref{thm:generalmethod} can be established in two stages.  First, it is straightforward to check by inspection of the proof of~\cite{DDS:15} that the proof of their Theorem~3.1 goes through essentially unchanged to establish the variant of the above theorem in which we substitute ``strong'' for ``weak'' in the third bullet.  (We note that the proof of~Theorem~3.1 of~\cite{DDS:15} requires a multiplicatively accurate evaluation oracle for each hypothesis distribution constructed in the course of the algorithm's execution, and that such oracles can be obtained in our setting since by our assumptions we have an approximate counting algorithm for the class $\class'$.) For the second stage, we claim that the variant of the above theorem with  ``strong'' in place of ``weak'' easily implies~\Cref{thm:generalmethod} as stated above (with ``weak'' in the third bullet).  To see this, observe that if $s$ draws are made from a distribution ${\cal D}'_{f,\tau}$ which has variation distance at most $\tau$ from ${\cal D}_f$, then the variation distance between the distribution of that sample of $s$ draws and the corresponding distribution (over samples of $s$ draws) from ${\calD}_f$ is at most $s\tau.$  Hence any procedure $A$ (such as the distribution learning algorithm $\Algo^{\class}$) which makes $s$ calls to a strong approximate generation algorithm can instead be run using a weak approximate generation algorithm, and if the $\tau$-parameter of the weak approximate generation algorithm is set to be $\delta/s$, this will decrease the success probability of the procedure $A$ by at most an additive $\delta.$ 

\medskip
\noindent {\bf Miscellaneous notation and terminology.} We write $B^1(z,r)$ to denote the $\ell_1$-ball of radius $r$ in $\R^n$ around a point $z$, and write $B^2(z,r)$ to denote the $\ell_2$-ball.

\section{A distribution learning algorithm for LTFs with respect to log-concave background distributions }\label{sec:upperbounds}

\subsection{Sampling and integration over log-concave distributions}
\label{sec:logconcave-sample-integrate}

The ability to efficiently sample from log-concave distributions, and to approximate various quantities such as the normalizing factor of a log-concave function, is crucial for  our upper bound result~\Cref{thm:pos-deg1-detailed}.  As mentioned in the introduction, there is an extensive literature on these problems. 
In this subsection we state and explain the specific results we shall rely on, and prove that our assumptions imply that we can indeed use them in our setting. In particular, the key take-away result from this subsection is the following crucial lemma:

\begin{lemma}~\label{lem:sampling-our-assump}
Let $\tilde{\mu}\colon \R^n \rightarrow [0,\infty)$ be a log-concave measure which is assumed to satisfy conditions (\textbf{C1}),  (\textbf{C2}) and (\textbf{C3}) from~\cref{sec:preliminaries}, and let $\mu \eqdef \tilde{\mu}/\normone{\tilde{\mu}}$ be the corresponding probability distribution. There is an algorithm with the following performance guarantee:  Given  access to an evaluation oracle for $\tilde{\mu}$, a halfspace $g\colon \R^n \rightarrow \{-1,1\}$, a confidence parameter $\delta$ 
and an error parameter $\eps> 1/2^{o(n)}$ such that $\Pr_{\bx \leftarrow \mu}[g(\bx)=1]\ge 2^{-n}$,  it runs in time $\poly(n,1/\eps, \log (1/\delta))$ and with probability $1-\delta$ outputs a number $\hat{p}$ such that 
\[
(1-\eps) \cdot \Pr_{\bx \leftarrow \mu}[g(\bx)=1] \leq \hat{p} \leq (1+\eps) \cdot \Pr_{\bx \leftarrow \mu}[g(\bx)=1].
\]
Let $\mu_{g}$ denote the distribution $\mu$ conditioned on $g(x)=1$. 
Under the above assumptions, there is also a randomized algorithm which runs in time $\mathsf{poly}(n,\log(1/\eps))$
and samples from a distribution $\nu$ such that $\Vert \nu - \mu_g\Vert_1 \le \eps$. 
\end{lemma}

\subsubsection{Setup for the proof of~\cref{lem:sampling-our-assump}}

We first observe that we can apply~\cref{thm:smooth} (see~\cref{ap:smooth}) to the log-concave measure $\tilde{\mu}$ to obtain a new measure $\tilde{\mu}_1$ which is a ``smoothed'' version of $\tilde{\mu}$.  As detailed in~\cref{thm:smooth}, the distribution $\mu_1$ corresponding to $\tilde{\mu}_1$ satisfies $\Vert \mu - \mu_1 \Vert_1 \le 2^{-3n/2}$, so $\mu$ and $\mu_{1}$ are statistically extremely close.  Let $\mu_g$ (resp. $\mu_{1g}$) denote the distribution $\mu$ (resp. $\mu_1$) conditioned on $g=1$. Since $\Pr_{\bx \leftarrow \mu} [g(\bx) =1] \ge 2^{-n}$, it follows that $\Vert \mu_g - \mu_{1g}\Vert_1 \le 2^{-n/2}$. Since we only work with error parameters $\eps \geq 1/2^{o(n)}$, the algorithms we consider only draw at most $2^{o(n)}$ many samples from $\mu_g$, and hence with very high probability sampling from $\mu_{1g}$ is identical to sampling from $\mu_g$. Thus, from now on, we will assume the following
(see~\cref{thm:smooth}):

\begin{claim}~\label{clm:extra-condition}
The log-concave measure $\tilde{\mu}$ satisfies the following conditions.
\begin{enumerate}
\item For $x \not \in B^2(0,n^C)$, $\tilde{\mu}(x) =0$. 
\item The resulting density $\mu$ is such that for all $x \in B^2(0,n^C)$, $\mu(x) \ge 2^{-2n^C -5n-3} \cdot n^{-2Cn}$. 
\end{enumerate}
\end{claim}

\cref{lem:sampling-our-assump} is proved by leveraging results of~\cite{LV06}. 
We first recall two assumptions from~\cite{LV06} which suffice for their results to go through.  Their algorithms deal with  an unknown log-concave function $\tilde{\mu}\colon\R^n\to[0,\infty)$ (and the associated probability density function $\mu\eqdef \tilde{\mu}/\normone{\tilde{\mu}}$), and are given

\begin{enumerate} 

  \item [\bf\small(LV1)] an evaluation oracle that returns the value of $\tilde{\mu}$ at any given $x\in\R^n$, as well as two values $R,r>0$ such that (i)~the variance of $\mu$ in every direction is at most $R^2$; and (ii)~if the level set
  $L(c) \eqdef \setOfSuchThat{ x\in\R^n }{\tilde{\mu}(x) \geq c }$ 
  satisfies $\mu(L(c)) \geq 1/8$, then $L(c)$ contains a ball of radius $r$; and
  
  \item [\bf\small(LV2)] a random ``starting point'' $\bx_0 \in \R^n$ drawn from a probability distribution $\sigma$, along with a bound $M\geq 1$ such  that $\frac{d\sigma}{d\mu} \leq M$.
  
\end{enumerate}

\begin{remark}
If condition {\bf\small(LV1)} holds for $R$ and $r$ such that $R/r = O(\sqrt{n})$, then the log-concave measure $\tilde{\mu}$ (and the resulting distribution $\mu$) is said to be \emph{well-rounded}. 
\end{remark}

Theorem~1.1 in \cite{LV06} states that under the conditions {\bf\small(LV1)} and {\bf\small(LV2)}, there is an efficient algorithm to approximately sample from the density $\mu$:

\begin{theorem}[Theorem~1.1, \cite{LV06}] \label{thm:LV-sampling}
There is an algorithm $A_{\text{LV-sampler}}$ which, under conditions {\bf\small(LV1)} and {\bf\small(LV2)}, given an error parameter $\eps >0$, has the following properties: 
\begin{enumerate}
\item The running time of $A_{\text{LV-sampler}}$ is $\poly(n, R/r, \log (1/\eps), \log M)$. 
\item The output distribution of the algorithm (denoted by $\nu$) satisfies $\Vert \mu - \nu \Vert_1 \le \eps$.   
\end{enumerate}
\end{theorem}

Note that the algorithm in~\Cref{thm:LV-sampling} runs in polynomial time provided that $\eps \geq 2^{-n^{O(1)}}$,
$M \leq 2^{n^{O(1)}}$, and $R/r \leq \poly(n).$ Moreover, when the distribution $\mu$ is well-rounded, then the next theorem (Theorem~1.3 in 
\cite{LV06}) also gives an efficient algorithm to compute $\int \tilde{\mu}(x)$. 
\begin{theorem}~\label{thm:LV-counting} 
There is an algorithm $A_{\text{LV-integral}}$ which, for a well-rounded log-concave function $\tilde{\mu}$, under conditions {\bf\small(LV1)} and {\bf\small(LV2)}, given access to an error parameter $\eps>0$, outputs a number $Q$ such that 
\[
1-\eps \le \frac{Q}{\int_x \tilde{\mu}(x) dx} \le 1+\eps. 
\]
The running time of the algorithm is $\poly(n/\eps)$. 
\end{theorem}

We stress that both~\cref{thm:LV-counting} and~\cref{thm:LV-sampling} rely on $R/r$ being not too large. While this 
 may not always be the case for the distribution $\mu$ to begin with, it is possible to achieve this via a linear transformation, as we discuss next. 

 \begin{definition}~\label{def:lin-trans-density}
 Let $T\colon \R^n \to \R^n$ be an invertible linear transformation, let $z \in \R^n$, and let $\mu$ be any measure  supported on $\R^n$. We define the measure $\mu_{T,z}$ which is given by 
 \[
 \mu_{T,z}(x) = \frac{1}{\det(T)} \mu(T^{-1}(x-z)). 
 \]
 \end{definition}
 We observe that if $\mu$ is a probability density, then $\mu_{T,z}$ is the density obtained by sampling $\bx \leftarrow \mu$ and applying the linear transformation $x \mapsto Tx + z$ to $\bx$.
Since linear transformations preserve log-concavity, if $\mu$ is log-concave then $\mu_{T,z}$ is also log-concave.

We further observe that for any log-concave measure $\tilde{\mu}$, any $z \in \R^n$, and any invertible matrix $T$, we have
\[
\int_{x} \tilde{\mu}(x) dx = \int_x \tilde{\mu}_{T,z}(x) dx
\]
Moreover, given an efficient algorithm that samples from a distribution that is $\eps$-close to the distribution $\mu_{T,z}$, there is also an efficient algorithm to sample from a distribution that is $\eps$-close to distribution $\mu$.

With the above observations in hand, a strategy to prove~\cref{lem:sampling-our-assump} naturally suggests itself. Given oracle access to a log-concave measure $\tilde{\mu}$ and a halfspace $g$ satisfying the conditions of~\cref{lem:sampling-our-assump}, we will compute a linear transformation $T\colon \R^n \rightarrow \R^n$ and a random starting point $\bz \in \R^n$ such that the density $\tilde{\mu}_{T,z}$ satisfies \textbf{LV1} with $R/r = O(\sqrt{n})$. In particular, we will prove the following lemma:

 \begin{lemma}~\label{lem:compute-isotropic-1}
 Let $\tilde{\mu}\colon \R^n \to [0,\infty)$ be a log-concave measure and $g$ be a halfspace satisfying the hypothesis of~\cref{lem:sampling-our-assump}. There is a polynomial time algorithm which, given evaluation oracle access to $\tilde{\mu}$ and a halfspace $g$, computes an invertible linear transformation $T\colon \R^n \to \R^n$ and a $z \in \R^n$ such that the density $({\mu}_{T,z})_g$ satisfies \textbf{LV1} with $R/r = O(\sqrt{n})$.
 \end{lemma}

\subsubsection{Proof of~\cref{lem:sampling-our-assump} assuming~\cref{lem:compute-isotropic-1}} \label{sec:proof-assuming}

 We now explain the details of how~\cref{lem:compute-isotropic-1} implies~\cref{lem:sampling-our-assump}. To do this, we need to recall the following result from~\cite{kannan1997random}:
 
\begin{theorem}[\cite{kannan1997random}]\label{theorem:sampling-convex}
Given a separation oracle for a convex body $K \subseteq \R^n$, there is an algorithm $A_{\text{KLS-sampler}}$ which makes $\tilde{O}(n^5)$ oracle calls and produces a uniformly random sample from $K$. 
\end{theorem}

 \begin{proofof}{\cref{lem:sampling-our-assump} assuming~\cref{lem:compute-isotropic-1}}
 Let us define the set $\mathcal{A}_g \subset \R^n$ to be ${\cal A}_g = \{x \in B^2(0,n^d) : g(x)=1\}$, and
 let us define $\sigma$ to be the uniform distribution over $\mathcal{A}_g$. We observe that
 we can efficiently sample from the distribution $\sigma$ because the set $\mathcal{A}_g$ has an separation oracle (recall that in the setting of~\cref{lem:sampling-our-assump} we are given $g$) and thus we can sample from $\mathcal{A}_g$ using the algorithm $A_{\text{KLS-sampler}}$ from~\cref{theorem:sampling-convex}. 
 Let $V$ denote the volume of the set $\mathcal{A}_g$, so $\sigma(x) = 1/V$ for all $x \in \mathcal{A}_g$. Since $\Pr_{\bx \leftarrow \mu}[g(\bx)=1] \ge 2^{-n}$ and $\mu(x) \le 2^n$ for all $x$ (by condition \textbf{C2}), it follows that $V \ge 2^{-2n}$, and consequently for all $x \in \mathcal{A}_g$, we have $\sigma(x) \le 2^{2n}$.
 
 By~\Cref{clm:extra-condition}, $\mu(x) \ge 2^{-2n^C-5n-3} \cdot n^{-2Cn}$ for every $x \in B^2(0,n^C)$. Since $\Pr_{\bx \leftarrow \mu} [g(\bx) =1] \ge 2^{-n}$, it follows that $\mu_g(x) \ge  2^{-2n^C-7n-3} \cdot n^{-2Cn}$. Consequently, we have that for every $x \in \mathcal{A}_g$, 
 \[
 \frac{\sigma(x)}{\mu_g(x)} \le 2^{2n^C + 9n +3} \cdot n^{2Cn}. 
 \]
 Hence for any invertible linear transformation $T\colon \R^n \rightarrow \R^n$ and any $z \in \R^n$, we also get
 \[
 \frac{\sigma_{T,z}(x)}{(\mu_{T,z})_g(x)} \le 2^{2n^C + 9n +3} \cdot n^{2Cn}. 
 \]
In particular, if we apply the linear transformation $T$ and $z$ by applying~\cref{lem:compute-isotropic-1} to the log-concave measure $\tilde{\mu}$ and the halfspace $g$, then we satisfy \textbf{LV1} and \textbf{LV2} with $R/r=O(\sqrt{n})$ and $M = 2^{2n^C + 9n +3} \cdot n^{2Cn}$, taking the point $\bx_0$ specified in {\bf LV2} to be a random point sampled from $\sigma$. We can now apply~\cref{thm:LV-counting} and~\cref{thm:LV-sampling} to obtain the desired conclusion of~\Cref{lem:sampling-our-assump}. 
 \end{proofof}

 \subsubsection{Proof of~\cref{lem:compute-isotropic-1}}
In the rest of this section we prove~\cref{lem:compute-isotropic-1}.  Towards this end, we first recall a useful definition:
\begin{definition}~\label{def:C-iso}
Let $\mu$ be a log-concave distribution supported on $\R^n$. For $\beta>0$, $\mu$ is said to be \emph{$\beta$-isotropic} if for every unit vector $u \in \R^n$, we have
\[
\frac{1}{\beta} \le \int_{x \in \R^n} \langle u,x \rangle^2 \mu(x) dx \le \beta. 
\]
In other words, the variance of $\mu$ in every direction is between $1/\beta$ and $\beta$. A log-concave measure $\tilde{\mu}$ is said to be $\beta$-isotropic if the corresponding distribution $\mu$ is $\beta$-isotropic.
\end{definition}
Lov\'asz and Vempala showed that if a log-concave distribution is $1$-isotropic, then it is well-rounded (Lemma~5.13, \cite{LV07}). 
However, it is easy to see by inspection of their proof that if a log-concave distribution is $\beta$-isotropic for any absolute constant $\beta>1$, then it is also well-rounded. We record this here: 
\begin{lemma}~\label{lem:well-rounded}
If a log-concave distribution $\mu$ is $\beta$-isotropic for an absolute constant $\beta>1$, then it is well-rounded. 
\end{lemma} 
We need two more ingredients to prove~\cref{lem:compute-isotropic-1}. The first is a result of Lov\'asz and Vempala which shows that given a maximizer of a log-concave measure, the measure can efficiently be brought to an $O(1)$-isotropic position. 

\begin{theorem} [\cite{LV07}] \label{thm:LV-isotropic}
Given evaluation oracle access to a full-dimensional log-concave measure $\tilde{\nu}$ and a point $y$ such that $\tilde{\nu}(y) \ge (1-2^{-n}) \cdot \max_x \tilde{\nu}(x)$, there is an algorithm which makes $\tilde{O}(n^4)$ oracle calls to $\tilde{\nu}$ and computes an invertible linear transformation $
T$ and a point $z\in \R^n$ such that the transformed measure $\tilde{\nu}_{T,z}$ (defined as in~\Cref{def:lin-trans-density}) is $2$-isotropic. 
\end{theorem}

  We remark that the theorem statement of~\Cref{thm:LV-isotropic} in \cite{LV07} stipulates that the point $y$ be an exact maximizer, as opposed to a near-maximizer; however, by inspecting the proof one can verify that having a near-maximizer (in the sense provided by our theorem statement) also suffices to find the linear transformation $T$ and point $z.$\footnote{This was confirmed in personal communication with Vempala~\cite{vempala-personal2018}.} 

The second ingredient we will use is the following theorem from~\cite{LSV17} (though weaker results might also suffice):
\begin{theorem}[{\cite[Theorem 1]{LSV17}}]~\label{thm:GLV-max}
Let $K$ be a convex set specified by a membership oracle, let $w \in \R^n$, and let $0<r'<R'$ be numbers such that $B^2(w, r') \subseteq K \subseteq B^2(w, R')$. For any convex function $f\colon \R^n \to \R$ given by an evaluation oracle and any $\eps>0$, there is a randomized algorithm that with constant probability computes a point $z \in B^2(K,\eps)$ such that 
\[
f(z) \le \min_{x \in K} f(x) + \eps \cdot \big(\max_{x \in K} f(x) - \min_{x \in K} f(x) \big) 
\]
 using $n^2 \log^{O(1)} \big(\frac{n \cdot R'}{\eps \cdot r'} \big)$ oracle calls and $n^3 \log^{O(1)} \big(\frac{n \cdot R'}{\eps \cdot r'} \big)$  arithmetic operations.
\end{theorem}

Now we are ready to prove~\Cref{lem:compute-isotropic-1}:
  
\begin{proofof}{\cref{lem:compute-isotropic-1}}
By applying~\cref{lem:well-rounded}, it suffices to compute $T,z$ such that the distribution $({\mu}_{T,z})_g$ is $2$-isotropic. We start by observing that $\tilde{\mu}_g(x)$ is proportional to $\tilde{\mu}(x) \cdot \mathbf{1}[g(x)=1]$, so in the setting of~\Cref{lem:compute-isotropic-1} we have evaluation oracle access to a measure proportional to $\tilde{\mu}_g$. By~\cref{thm:LV-isotropic}, it suffices to find a point $y \in \R^n$ such that $\tilde{\mu}_g(y) \ge (1-2^{-n}) \cdot \max_{x \in \mathcal{A}_g} \tilde{\mu}_g(x)$. 

To find the required point $y$, we will use~\cref{thm:GLV-max}. Note that since $\tilde{\mu}_g$ is log-concave, the function $-\log \tilde{\mu}_g$ is convex, so minimizing $h(x) = -\log \tilde{\mu}_g(x)$ is equivalent to maximizing $\tilde{\mu}_g(x)$. We now provide the details.  Recalling that $\|\mu\|_\infty \leq 2^n$ and $\Pr_{\bx \leftarrow \mu}[g(\bx)=1] \geq 2^{-n}$, we get that $\|\mu_g\|_\infty \leq 2^{2n}$; combining this with the lower bound on $\mu_g(x)$ established in~\Cref{sec:proof-assuming}, we get that
\[
2^{-2n^C-7n-3} \cdot n^{-2Cn} \le {\mu}_g(x) \le 2^{2n}
\]
for all $x \in B^2(0,n^C)$.
We take the set $K$ of~\cref{thm:GLV-max} to be $K = \mathcal{A}_g$, and we observe that by the above bounds on $\mu_g$, we have
\begin{equation}~\label{eq:diff-log}
\max_{x \in K} h(x) - \min_{x \in K} h(x) \le 2n^C + 2Cn \log n + 9n +3. 
\end{equation}
We further observe that we clearly have both a separation oracle and a membership oracle for the set $K$.  We use this separation oracle with~\cref{theorem:sampling-convex} to generate a point $\bw$ which is uniformly sampled from $K$. We will now show that setting
\[
R'= 2n^C, \ \ \ \ r' = 2 \cdot (2n)^{-(n-1)C} \cdot \pi^{-(n-1)} \cdot 2^{-3n},
\]
we have $B^2(\bw,r') \subseteq K \subseteq B^2(\bw,R')$. 
The second inclusion is clear (that for the above $R'$, we have $K \subseteq B^2(w,R')$). For the first inclusion we recall the following fact from high-dimensional convex geometry (which follows from the existence of higher-dimensional Crofton formulas, see, e.g., \cite{surface-area-enclosed}):
 
\begin{fact}~\label{fact:convex-area}
Let $K$ be any convex body contained in a ball of radius $\rho = n^C$. Then the surface area of $K$ is at most the surface area of the sphere of radius $\rho$ (which is bounded by $2  \pi^{n-1} \rho^{n-1}$). 
\end{fact}

We will now use this to bound the volume of the set of the points at a distance $r'$ from the boundary of $\mathcal{A}_g$. To do this, for any convex body $K$, define $K_t = \{x: \mathsf{dist}(x, K) \le t\}$. Observe that $K_t$ is a convex body for any $K$ and $t$. The following relation is well-known: 
\[
\mathsf{surf}(K) = \lim_{\delta \rightarrow 0} \frac{\mathsf{vol}(K_\delta) - \mathsf{vol}(K)}{\delta}. 
\]
Here $\mathsf{surf}(K)$ is the surface area of $K$. Thus, it follows that for any $t \ge 0$, 
\begin{equation}~\label{eq:vol-integral}
\mathsf{vol}(K_t) - \mathsf{vol}(K) = \int_{s=0}^t  \mathsf{surf}(K_s) ds. 
\end{equation} 
Let $A_{g,t} =  \{x: \mathsf{dist}(x, A_g) \le t\}$. Applying Fact~\ref{fact:convex-area} to $A_{g,t}$, we have $\mathsf{surf}(A_{g,t}) \le 2 \pi^{n-1} \cdot (n^C +t)^{n-1}$. This is because if $A_g$ is contained in a ball of radius $n^C$, then  $A_{g,t}$ is contained in a ball of radius $n^C+t$. Combining this 
with (\ref{eq:vol-integral}) applied to the  body $A_g$, 
\[
\mathsf{vol}(A_{g,r'}) - \mathsf{vol}(A_{g}) \le \int_{s=0}^{r'} 2 \pi^{n-1} \cdot (n^C +s)^{n-1} ds
\]
Using the fact that $r' \le n^C$, we have that $\mathsf{vol}(A_{g,r'}) - \mathsf{vol}(A_{g}) \le 2  \pi^{n-1} (2n)^{(n-1)C} \cdot r'$. 
Thus, the total volume of the set of points at a distance $r'$ from the boundary of $\mathcal{A}_g$ is at most $2  \pi^{n-1} (2n)^{(n-1)C} \cdot r'$. 
By plugging in the value of $r'$ from above, this volume is at most $2^{-3n}$.  Since $\mathsf{Vol}(\mathcal{A}_g) \ge 2^{-2n}$ (see ~\Cref{sec:proof-assuming}), this means that with probability $1-2^{-n}$, the point $\bw$ satisfies 
$B^2(w,r') \subseteq K \subseteq B^2(w,R')$. 

\noindent We now set $\eps$ to be 
\[
\eps \eqdef \frac{2^{-2n-1}}{2n^C + 2Cn \log n + 9n +3}. 
\]
We apply~\cref{thm:GLV-max} with $w$, $R'$, $r'$ and $\eps$ as above 
to get a point $z$ such that 
\[
h(z) \le \min_{x \in K} h(z) + \eps \cdot \big(\max_{x \in K} h(x) - \min_{x \in K} h(x) \big) \le \min_{x \in K} h(x) + 2^{-2n-1}  \ \textrm{ using~\eqref{eq:diff-log}.}
\]
Thus, $\tilde{\mu}_g(z) \ge e^{-2^{-2n-1}} \cdot\max_{x \in K} \tilde{\mu}_g(x) $, which implies that $\tilde{\mu}_g(z) \ge (1-2^{-n}) \max_{x \in K} \tilde{\mu}_g(x)$. With this point $z$ in hand, we can apply~\cref{thm:LV-isotropic} to bring $\tilde{\mu}_g$ to $2$-isotropic position. This finishes the proof. 
\end{proofof}

  \subsection{An algorithm for LTFs under well-behaved log-concave distributions}   \label{sec:upperbounds:1ptf:logconcave}  
  
In this subsection we prove~\cref{thm:pos-deg1-detailed}.  We do this by applying~\cref{thm:generalmethod}, where both the classes $\class$ and $\class'$ are taken to be the class $\ptf{n}{1}$ of linear threshold functions over $\R^n$.  Thus what is needed is to establish the existence of each of the four algorithmic ingredients --- a densifier, an approximate counting algorithm, a weak approximate generation algorithm, and an SQ-learning algorithm --- required by~\Cref{thm:generalmethod}.  (Some of these will be reused in the proof of~\Cref{thm:pos-deg2-detailed} as well.)

Several of these ingredients are provided for us in the literature.  An SQ-learning algorithm for the class of linear threshold functions over $\R^n$ was given in \cite{BFK+:97}:
\begin{theorem}~\label{thm:BFKV}
There is a distribution-independent {\rm SQ} learning algorithm $\Algo_{\rm SQ}^{\ltf}$ for the class of linear threshold functions over $\R^n$ that has running time $\poly(n,1/\eps, \log(1/\delta))$, uses at most $\poly(n)$ time to evaluate each query and requires tolerance of its queries no more than $\tau = 1/\poly(n,1/\eps)$. 
\end{theorem}

An approximate counting algorithm for linear threshold functions under well-behaved log-concave distributions is given by~\Cref{lem:sampling-our-assump}, and this lemma also gives a weak approximate generation algorithm.  Thus, it remains only to establish the existence of an efficient densifier for linear threshold functions under log-concave background distributions.

\medskip
\noindent 
{\bf Construction of a densifier for $\ptf{n}{1}$ under well-behaved log-concave distributions.}
To explain the construction of the densifier, we will need to recall the notion of an online learning algorithm. An online algorithm $\mathcal{A}$ for a concept class $\mathcal{C}$ operates in a sequential model of learning: at every stage, the algorithm is presented with an unlabeled point $x$ and is asked to predict the binary value $f(x)$ where $f$ is the target function. After the prediction, the true value of $f(x)$ is revealed. $\mathcal{A}$ is said to have a mistake bound of $M$ if the maximum number of incorrect predictions made by the algorithm is $M$ on any sequence of unlabeled examples and any target function $f \in \mathcal{C}$. 

We will use the following celebrated result from~\cite{MaassTuran:94}, which gives an online learning algorithm 
for $\ptf{n}{1}$ if the example points come from a discrete grid:
\begin{theorem}~\label{thm:Maass-Turan}
There is an online learning algorithm (which we call $\mathcal{M}\mathcal{T}_{n,L}$) for $\ptf{1}{n}$ over the domain $[L]^n$   with the following guarantee:  $\mathcal{M}\mathcal{T}_{n,L}$ has a mistake bound of $M(L,n) \eqdef O(n^2 (\log L + \log n))$ and runs in time $\poly(n,\log L)$ at each stage. Further, at every stage of its execution, $\mathcal{M}\mathcal{T}_{n,L}$ maintains a (weights based) representation of a LTF which correctly labels all the examples received so far. 
\end{theorem}
We now state our main theorem establishing the existence of densifiers for $\ptf{1}{n}$ when the ambient distribution is a well-behaved log-concave distribution:

\begin{theorem}~\label{thm:densifier-log-concave}
Let $\mathcal{D}$ be any log-concave density satisfying conditions \textbf{C1}, \textbf{C2} and \textbf{C3}. There is an algorithm $\mathcal{A}^{\ltf}_{\mathsf{den}}$ which is an $(\eps, \gamma, \delta)$-densifier for $\ptf{n}{1}$ using $\ptf{n}{1}$
under $\mathcal{D}$ 
 where $\gamma = 1/\mathsf{poly}(n).$  Algorithm $\mathcal{A}^{\ltf}_{\mathsf{den}}$  has running time $\mathsf{poly}(n,1/\eps, \log(1/\delta))$. 
\end{theorem}
The proof of~\cref{thm:densifier-log-concave} is a modification of the analogous proof in \cite{DDS:15} (Theorem~11 in the full version) where the background distribution $\mathcal{D}$ is uniform over $\{-1,1\}^n$, so, instead of giving a full proof, we limit ourselves to a proof sketch. (The key new issue that needs to be dealt with is the fact that mistake-bound guarantees do not hold in a completely unrestricted setting of learning over continuous domains.  At a high level we surmount this issue by exploiting specific properties of our well-behaved log-concave distributions, such as the fact that such a density cannot be too concentrated in a very small region.)

\subsubsection{Proof sketch of~\cref{thm:densifier-log-concave}} \label{sec:sketches}
 We begin by recalling that to show a function $g$ is a densifier for $f$, it suffices to establish two properties: 
\begin{description}
  \item[(a)] Agreement:  $\probaDistrOf{\bx\leftarrow \calD_f}{g(\bx)=1} \geq 1- \eps$, and
  \item[(b)] Density:  $\probaDistrOf{\bx\leftarrow \calD}{g(\bx)=1} \leq (1/\gamma) \probaDistrOf{\bx\leftarrow \calD}{f(\bx)=1}$.
\end{description} 

\medskip
\noindent {\bf A densifier for LTFs  in the uniform-distribution \cite{DDS:15} setting.}
We first recall the algorithm to construct  a densifier from \cite{DDS:15} and give a high level overview of its proof of correctness (Theorem~11 in \cite{DDS:15}). Note that \cite{DDS:15} constructs a densifier for $\ptf{n}{1}$ where the domain is $\{-1,1\}^n$ (or equivalently, $[2]^n$) and the background distribution $\mathcal{D}$ is uniform on $\{-1,1\}^n$. In the following, the value of $\gamma$ is chosen to be a suitably small $1/\poly(n)$ quantity (small compared to $1/M$ as defined below).

\begin{enumerate}
\item Sample $N_+ = \Theta((n^2+\log(1/\delta))/\eps^2)$ points $\bx_1, \ldots, \bx_{N_+}$ where each $\bx_i$ is drawn from 
${\cal D}_f$ (i.e.,~${\cal D}$ conditioned on $f(\bx_i)=1$).
\item Define the set $\mathcal{S}^+ \eqdef \{\bx_1, \ldots, \bx_{N_+}\}$. Let  $M = O(n^2 \log n)$ and $i=0$. 
\item Initialize the online algorithm $\mathcal{M}\mathcal{T}_{n,L}$ with $L=2$ and let $h_0$ be the initial hypothesis. 
\item If $i=M$, output $h_i$ and terminate. 
\item  If there exists $x \in S^+$ such that $h_i(x) =-1$, then feed such an $x$ to $\mathcal{M}\mathcal{T}_{n,L}$ along with the label $\ell_x=1$ for $x$. 
Go to Step~4. 
\item Otherwise, use the approximate counting algorithm to get $\widehat{p}_i$, a high-accuracy approximation of
$\Pr_{\bx \leftarrow {\cal D}}[h_i(\bx)=1]$.
\item If $\widehat{p} \geq \frac{\gamma}{2} \cdot \widehat{p}_i$,  then terminate the procedure and output $h_i$. 
\item Otherwise,  generate $\bx \leftarrow \mathcal{D}_{h_i}$ and feed $\bx$ to $\mathcal{M}\mathcal{T}_{n,L}$ along with the label $\ell_{\bx}=-1$ for $\bx$. Update hypothesis $h_i$ using $\mathcal{M}\mathcal{T}_{n,L}$ and set $i \leftarrow i+1$.  Go to Step~4.
\end{enumerate}
The proof of correctness of the algorithm in \cite{DDS:15} relies on two claims. 
\begin{claim}~\label{clm:correctness-1}
With very high probability over the draw of ${\cal S}^+$, (a) is satisfied by any halfspace $g\colon \{-1,1\}^n \to \{-1,1\}$ such that  $g(x)=1$ for every $x \in S^+$.
\end{claim}
\begin{claim}~\label{clm:correctness-2}
With very high probability over the execution of the densifier, every sample $x$ and label $\ell_x$ fed to $\mathcal{M}\mathcal{T}_{n,L}$ is \emph{correct}, i.e., $f(x) = \ell_x$. 
\end{claim}

(For intuition,~\cref{clm:correctness-1} is argued using a simple Chernoff bound and union bound.~\cref{clm:correctness-2} exploits the fact that since $\gamma$ is very small compared to $1/M$ where $M$ is the number of stages, with high probability each random point $\bx \leftarrow {\cal D}_{h_i}$ obtained in Step~8 is indeed a negative example for $f$.)
Given~\cref{clm:correctness-2} and~\cref{clm:correctness-1}, it is not hard to show that with high probability, the output of the above algorithm terminates with a hypothesis $h_i$ which satisfies  both (a) and (b) and thus we have a densifier.

\medskip
\noindent {\bf Adapting the \cite{DDS:15} densifier to our setting.}
We now return to our setting in which the background distribution $\mathcal{D}$ is log-concave and satisfies \textbf{C1}, \textbf{C2} and \textbf{C3}.  We change the densifier algorithm as described below.

Let $\kappa = 2^{-n^{10 C}}$ and let $L = 1/\kappa$. For any point $x \in \mathbb{R}^n$, let $[x]_\kappa$ denote the point obtained by rounding each coordinate of $x$ to the nearest integral multiple of $\kappa$, and let $M = O(n^2 (\log L + \log n))$ (note that this new value of $M$ is still $\poly(n)$). We adapt the algorithm of \cite{DDS:15} to our setting in the following way: 
\begin{enumerate}
\item Set $L$ and $M$ in the algorithm with the values specified above. 
\item The background distribution is the log-concave distribution $\mathcal{D}$ satisfying \textbf{C1}, \textbf{C2} and \textbf{C3}. 
\item For each point $x$ fed to the learner in Steps~5 and~8, we instead feed $[x]_\kappa$ to the learner (so each point fed to the learner is rounded to the nearest integral multiple of $\kappa$).  
\end{enumerate}

These changes induce analogous changes to~\cref{clm:correctness-1,clm:correctness-2}, where now the points $[x]_{\kappa}$ play the role that $x$ played throughout the statements; more precisely, we state

\medskip

\noindent {\bf New Version of~\cref{clm:correctness-1}.} \emph{With very high probability over the draw of ${\cal S}^+$, the bound $\Pr_{\bx \leftarrow {\cal D}_f}[g([\bx]_\kappa)=1] \geq 1 - 2 \eps$ is satisfied by any halfspace $g\colon [L]^n \rightarrow \{-1,1\}$ such that  $g(x)=1$ for every $x \in S^+$.
}
\medskip

\noindent {\bf New Version of~\cref{clm:correctness-2}.} \emph{With very high probability over the execution of the densifier, every sample $[x]_\kappa$ and label $\ell_x$ fed to $\mathcal{M}\mathcal{T}_{n,L}$ is \emph{correct}, i.e., $f([x]_\kappa) = \ell_x$. 
}

\medskip

We now give proof sketches of these new versions of~\cref{clm:correctness-1,clm:correctness-2} in our context.  (The high-level idea throughout the arguments given below is that since the rounded version $[x]_\kappa$ is always extremely close to $x$ and the relevant distributions are not too concentrated, they can only put a tiny amount of mass on points close to the boundary, and thus it is very unlikely that the rounding causes the label to change for any point arising in the algorithm.)

\begin{proofsketchof}{New Version of~\cref{clm:correctness-1}}
Let $g$ be any halfspace and define the set $K_g \eqdef g^{-1} (-1) \cap f^{-1}(1)$. Observe that $K_g$ is an intersection of two LTFs and further that if $g$ does not satisfy (a), then
\begin{equation} \label{eq:goodie}
\Pr_{\bx \leftarrow \mathcal{D}_f} [\bx \in K_g] = \Pr_{\bx \leftarrow \mathcal{D}_f} [\bx \in g^{-1}(-1)] \ge \eps.
\end{equation}

We upper bound the probability  
$\Pr_{\bx \leftarrow \mathcal{D}_f} [\bx \in K_g \textrm{ and } [\bx]_\kappa \not \in K_g]$. To bound this, observe that since $\mathcal{D}$ satisfies \textbf{C2}, instead of $K_g$, we can consider $K_g \cap B^2(0,n^C)$. Since the surface area of this (convex) body is upper bounded by $n^{C(n-1)} \cdot\pi^{n-1} \cdot 2^n$  by \cref{fact:convex-area}, the surface area of the portion of the two hyperplanes defined by $f$ and $g$ which is within $B^2(0,n^C)$ is at most $n^{C(n-1)} \cdot\pi^{n-1} \cdot 2^n$. We additionally recall that the density of $\mathcal{D}_f$ is bounded by $4^n$. Now, the event 
``$\bx \in K_g \textrm{ and } [\bx]_\kappa \not \in K_g$'' implies that $\bx$ lies in $B^2(0,n^C)$ and is $\kappa$-close to a point on the separating hyperplane for the LTF $f$ or a point on the separating hyperplane for the LTF $g$. By the above surface area and density upper bounds, this probability can be upper bounded by 
\[
n^{C(n-1)} \cdot\pi^{n-1} \cdot 2^n \cdot \kappa \cdot 4^n \ll \frac{\eps}{2}
\]
(recall that by assumption the value of $\eps$ is at least $1/2^{o(n)}$).
Combining this with \Cref{eq:goodie}, we conclude that $\Pr_{\bx \leftarrow \mathcal{D}_f} [[\bx]_\kappa \in g^{-1}(-1)] \ge \eps/2$.  Since $g$ is a halfspace and the VC-dimension of halfspaces is $n+1$, a standard uniform convergence argument now gives the claim. 
\end{proofsketchof}

\begin{proofsketchof}{New Version of~\cref{clm:correctness-2}}
Consider any example $(x,\ell_x)$ fed to the online learner. We first consider the case when $\ell_x=1$ (corresponding to Step~5). In this case, note that $x=[\by]_\kappa$ and $\ell_x=f(\by)=1$ where $\by \leftarrow \mathcal{D}_{f}$. Similar to before, note that the only way $\ell_x$ can differ from $f(x)$ is if 
$\by \in B^2(0,N^c)$ and $\by$ is $\kappa$-close to the boundary of the set $K$, where now $K = f^{-1}{(1)}$. As in the argument in the previous proof sketch, this probability can be bounded by 
\[
n^{C(n-1)} \cdot\pi^{n-1} \cdot 2^n \cdot \kappa \cdot 4^n \ll 2^{-n}. 
\]
Next we consider the case when $\ell_x=-1$ (corresponding to Step~8). In this case, we note that the point $\bx$ generated in Step~8 is indeed in $f^{-1}(-1)$ with probability at least $1-\gamma$. Consequently, $\Pr [[\bx]_\kappa \not \in f^{-1}(-1)] \le \gamma + 2^{-n}$. Since the total number of points the algorithm generates in Step~8 is at most $M$, the claim follows from the fact that $M(\gamma + 2^{-n}) \ll 1$ (as in the analysis of \cite{DDS:15}).
\end{proofsketchof}

Given the new versions of~\cref{clm:correctness-1,clm:correctness-2}, an argument entirely analogous to the argument employed in \cite{DDS:15} to establish {\bf (a)} and {\bf (b)} from~\cref{clm:correctness-1,clm:correctness-2} gives us the following analogues of {\bf (a)} and {\bf (b)}:

\begin{description}
  \item[(a$'$)] Agreement in our setting:  $\probaDistrOf{\bx\leftarrow \calD_f}{g([\bx]_{\kappa})=1}\geq 1- 2\eps$,
  \item[(b$'$)] Density in our setting:  $\probaDistrOf{\bx\leftarrow \calD}{g([\bx]_\kappa)=1} \leq (1/\gamma) \probaDistrOf{\bx\leftarrow \calD}{f(\bx)=1}$.
\end{description} 

Similar arguments to those employed above show that the probability (under either ${\cal D}_f$ or ${\cal D}$) that $\bx$ is so close to the boundary of $g$ as to have $g([\bx]_\kappa) \neq g(\bx)$ is extremely small.  Hence conditions {\bf (a)} and {\bf (b)} can be inferred from {\bf (a$'$)} and {\bf (b$'$)}, as required by the definition of a densifier.
This concludes the proof sketch of~\cref{thm:densifier-log-concave}.

  \section{A distribution learning algorithm for degree-2 PTFs under the normal background distribution}\label{sec:upperbounds:2ptf:gaussian}

In this section we fix the ambient distribution to be a standard Gaussian, $\calN\eqdef \calN(0,1)^n$. We wish to learn the distribution ${\cal N}_f$ induced by $\calN$ and an (unknown) degree-2 PTF $f$. In this section, we give an algorithm with time and sample complexity $\poly(n,1/\eps)$ which learns ${\cal N}_f$ to total variation $\eps$, given independent samples from it.

\begin{theorem}[{\Cref{thm:pos-deg2-detailed}, restated}]\label{theo:ub:2ptf:gaussian}
There is an algorithm $A$ with the following property:  Let $f\in\ptf{n}{2}$ be such that $\Pr_{\bx \gets {\cal N}(0,1)^n}[f(\bx)=1] \geq 2^{-n}.$  Given access to i.i.d.~samples from $\calN_f =$ ($\calN(0,1)^n$ restricted to $f^{-1}(1)$), for $\eps = 1/2^{o(n)}$ algorithm $A$ runs in $\poly(n,1/\eps)$ time and with high probability outputs a hypothesis distribution $\calD'$ such that $\dtv(\calD',\calN_f) \leq \eps.$
\end{theorem}

As in the previous section, to prove this theorem we will rely on the general algorithmic result of~\Cref{thm:generalmethod}, with $\class$ and $\class'$ both set to be $\ptf{n}{2}$ and $\calD$ being  $\calN(0,1)^n$. That is, we will provide (i)~a densifier (\Cref{sec:upperbounds:2ptf:gaussian:densifier}); (ii)~an approximate counting algorithm (\Cref{sec:upperbounds:2ptf:gaussian:counting}); (iii)~a weak approximate generation algorithm (\Cref{sec:upperbounds:2ptf:gaussian:sampling}); and (iv)~an SQ learning algorithm for $\ptf{n}{2}$.\footnote{We note that for (ii) and (iii), we will actually provide a marginally less general result, which we will show to be sufficient for our purposes given the assumptions of~\Cref{theo:ub:2ptf:gaussian}: namely, counting and generation algorithms only for $f \in \ptf{n}{2}$ such that $\Pr_{\bx \gets {\cal N}(0,1)^n}[f(\bx)=1] \geq 2^{-n}$.}
In fact, (iv) follows from~\Cref{thm:BFKV}, since upon reparameterization a degree-2 PTF over $\R^n$ is a degree-1 PTF over $\R^m$, for $m\eqdef n + \binom{n}{2}$.

\subsection{Construction of densifier for $\ptf{n}{2}$ under normal distribution}\label{sec:upperbounds:2ptf:gaussian:densifier}
In this subsection, we establish the following theorem, which provides a densifier for $\ptf{n}{2}$ under the ambient normal distribution:
\begin{theorem}~\label{thm:densifier-normal}
There is an algorithm $\mathcal{A}^{\sf PTF}_{\mathsf{den}}$ which is an $(\eps, \gamma, \delta)$-densifier for $\ptf{n}{2}$ using $\ptf{n}{2}$ under $\calN$ 
 where $\gamma = 1/\mathsf{poly}(n).$  Algorithm $\mathcal{A}^{\sf PTF}_{\mathsf{den}}$  has running time $\mathsf{poly}(n,1/\eps, \log(1/\delta))$. 
\end{theorem}
\begin{proof}[Proof sketch]
We follow the same outline as for the proof sketch of~\Cref{thm:densifier-log-concave}. Indeed, the algorithm itself is, up to some change in parameters, the same: we will again rely on the online-mistake bound learning algorithm of Maass--Turan from \Cref{thm:Maass-Turan} (observe that the Maass--Turan algorithm can be used to learn degree-2 PTFs over a discrete domain $[L]^n$, by viewing such a PTF as an LTF over $O(n^2)$ variables), and we will use the same approach as in the previous section.

To make this approach work for degree-2 PTFs under the normal distribution, it suffices to (1) adapt the algorithm in two places (namely, by using the approximate counting and approximate generation algorithms for degree-2 PTFs, given in~\Cref{sec:upperbounds:2ptf:gaussian:counting,sec:upperbounds:2ptf:gaussian:sampling}, instead of their LTF counterparts); and (2) establish the analogues of Claims~\ref{clm:correctness-1} and~\ref{clm:correctness-2} in this new setting (which will correspond to slightly different choices for the parameters $N_+$ and $M$).

The main hurdle in doing this is that the proofs of Claims~\ref{clm:correctness-1} and~\ref{clm:correctness-2}, in order to establish that the discretization of the space and rounding of the points (required to apply the Maass--Turan algorithm over a discrete grid of points) did not affect correctness, used the fact that the relevant sets (e.g., $g^{-1}(-1)\cap f^{-1}(1)$ and $f^{-1}(1)$) were convex sets with a very simple structure (the satisfying assignments of a single LTF or an intersection of two LTFs).  This simple structure made it possible to bound the volume of the set of points in $B^2(0,n^C)$ that are very close to the boundary of these sets using Fact~\ref{fact:convex-area}. While the rest of the argument goes through for degree-2 PTFs (using, for the uniform convergence in the analogue of Claim~\ref{clm:correctness-1}, that the VC dimension of the class of degree-$2$ PTFs over $n$ real variables is $\binom{n+2}{2}$~\cite{Anthony:95}), we cannot bound the probability that the rounding affects the value of the function in the same way as before since we are no longer dealing with LTFs; indeed, the relevant sets of satisfying assignments need no longer be convex.

Instead, we bound this probability more directly by exploiting the fact that the ambient distribution is Gaussian, as follows. First, we may assume without loss of generality that the current hypothesis degree-2 PTF $g = \sign(p)$ has been normalized, so that the squared non-constant coefficients of $p$  sum to 1. Given any $x\in B^2(0,n^C)$, consider its rounding $[x]_\kappa$. Using Cauchy--Schwarz, the triangle inequality, and the fact that the coefficients of $p$ have been normalized as described above, it is not hard to see that 
\[
    \abs{ p(x) - p([x]_\kappa) } \leq O(1)\cdot \kappa\cdot \normtwo{x} = O(\kappa n^C) \eqdef \tau\,.
\]
Consequently, one can only have $g(x) \neq g([x]_\kappa)$ if $p(x)$ itself is very small, namely at most $O(\kappa n^C)$. However, invoking a well-known result of Carbery and Wright~\cite{CW:01}, this itself only can happen with probability $O(\sqrt{\tau}) = O(\sqrt{\kappa n^C})$ under the standard normal distribution:
\begin{theorem}[\cite{CW:01}]~\label{thm:CW}
  Let $p\colon \R^n\to \R$ be a non-identically zero degree-$d$ polynomial. Then, for all $\eps >0$ and $\theta\in\R$,
  \[
   \Prx_{\bx\leftarrow\calN(0,1)^n} \left[ \abs{p(\bx)-\theta} < \eps\sqrt{\Var[ p]} \right] \leq O(d \eps^{1/d})\,.
  \]
\end{theorem}
Since $p$ is degree-2, $\Var[p]$ is the sum of the squared degree-1 and degree-2 Hermite coefficients when $p$ is expressed in the Hermite basis of polynomials that are orthonormal under ${\cal N}(0,1)^n$.  Recalling that the degree-1 univariate Hermite polynomial is $h_1(x)=x$ and the degree-2 univariate Hermite polynomial is $h_2(x) = {\frac {x^2 - 1}{\sqrt{2}}}$, the fact that $p$'s squared non-constant coefficients sum to 1 implies that $\Var[p] = \Theta(1).$ In our setting $\theta=0$, and $d=2$, and so applying \Cref{thm:CW} we get the claimed bound.

 We can use this to derive the counterparts of the main probability bounds from Claim~\ref{clm:correctness-1} and Claim~\ref{clm:correctness-2}, respectively, allowing the proofs of the two analogous statements to go through. First, letting $K_g \eqdef g^{-1}(1)\cap f^{-1}(-1)$, we have
\[
    \Prx_{\bx\leftarrow \calN_f}[\bx \in K_g \cap [\bx]_\kappa \notin K_g ]
    \leq  \Prx_{\bx\leftarrow \calN_f}[\abs{p(\bx)} \leq \tau ]
    \leq  2^n \Prx_{\bx\leftarrow \calN}[\abs{p(\bx)} \leq \tau ]
    \leq O(2^n \sqrt{\tau}) \ll 2^{-n}
\]
recalling that $\probaDistrOf{\bx\leftarrow \calN}{ f(\bx) = 1 } \geq 2^{-n}$ and the setting of $\kappa$. This (along with the aforementioned VC dimension argument) yields the counterpart of Claim~\ref{clm:correctness-1}.  A similar argument applied to the degree-2 PTF $f$ (instead of $g$) allows us, considering $K \eqdef  f^{-1}(-1)$ and following the outline of the proof of  Claim~\ref{clm:correctness-2}, to establish the counterpart of that second claim as well.
\end{proof}

\subsection{Approximate counting: Gaussian integration under quadratic constraints}\label{sec:upperbounds:2ptf:gaussian:counting}

Recall that our goal in this subsection is an algorithm with the following property: given the full description of a degree-2 PTF $g=\sign(p)$ over $\R^n$ such that 
$g^{-1}(1)$ has mass at least $1/2^n$ under $\calN$, the algorithm should efficiently output a (multiplicatively accurate) estimate of $\Prx_{\bG \sim \calN}[g(\bG)=1].$ 
The main theorem of this subsection is the following.

\begin{theorem}~\label{thm:PTF-counting}
   There is a deterministic algorithm $\mathcal{AC}$ with the following guarantee: Given as input a degree-2 polynomial $p$ over $\R^n$ such that $g^{-1}(1)$ has mass at least $1/2^n$ under $\calN$ (where $g=\sign(p)$) and input $\eps\in(0,1]$ such that $\eps \geq 1/2^{O(n)}$, $\mathcal{AC}$ runs in time $\poly(n,1/\eps)$ and outputs an $(1+\eps)$-approximation to $\probaDistrOf{\bG\leftarrow\calN}{g(\bG) = 1}$.
\end{theorem}

Although sampling $\bG=(\bG_1,\dots,\bG_n)\leftarrow \calN$ is easy, the additional quadratic constraint significantly complicates the problem of both  (i) estimating $\probaOf{p(\bG) \geq 0}$ as well as (ii) sampling $\bG$ conditioned on $\probaOf{p(\bG) \geq 0}$. Indeed, as is the case with many sampling and approximate counting problems~\cite{JVV86}, the complexities of (i) and (ii) are closely related.  We note that these problems are of interest in the statistics literature~\cite{pakman2014exact}, but rigorous guarantees were previously not known. 

To prove \Cref{thm:PTF-counting}, we leverage an approximate counting algorithm due to Li and Shi~\cite{LS:14} for the knapsack problem (which in turn builds on the algorithm of \cite{vstefankovivc2012deterministic}). To state the result of Li and Shi, we first need the definition of an \emph{oracle to a probability distribution}.

\begin{definition}
For any real valued random variable $\bX$, an oracle $\mathcal{O}_{\bX}$ takes inputs $\nu_1 \le \nu_2 \in \mathbb{R}$ and outputs the quantity  $\Pr[\nu_1 \le \bX \le \nu_2]$. 
\end{definition} 
With this, we are ready to state the result of Li and Shi. 
\begin{theorem}[{\cite[Theorem 1.1]{LS:14}}] \label{theo:lishi}
  There exists an algorithm $\mathcal{LS}$ with the following guarantees. Given as input an integer $C>0$, an approximation parameter $\eps > 0$, and oracle access to $n$ independent integer-valued random variables $\bX_1,\dots,\bX_n$ such that $0\in \operatorname{supp}(\bX_i) \subseteq \{0,1,\dots, C+1\}$ for all $1\leq i\leq n$, the algorithm outputs a value $\hat{\rho}$ such that
  \[
          \frac{1}{1+\eps} \probaOf{ \sum_{i=1}^n \bX_i \leq C } \leq \hat{\rho} \leq (1+\eps) \probaOf{ \sum_{i=1}^n \bX_i \leq C }\,.
  \]  
  Moreover, $\mathcal{LS}$  runs in time $\poly(n,\frac{1}{\eps},\log C,\log \frac{1}{\Delta})$, where $\Delta \eqdef \prod_{i=1}^n \probaOf{ \bX_i = 0}$.
\end{theorem}

To see why \Cref{theo:lishi} is useful, we begin by observing that, upon rotating $\R^n$ and translating the Gaussians, the approximate counting problem considered in \Cref{thm:PTF-counting} is equivalent to the following problem:
\begin{framed}
Given $(\lambda_i,\mu_i)_{i\in [n]}$, $\eps\in(0,1]$, and  $\theta\in\R$ such that (\ref{eq:problem:counting}) is at least $2^{-n}$,  output an $(1\pm\eps)$-approximation of
\begin{equation}\label{eq:problem:counting}
\Prx_{\bG \sim {\cal N}}[p(\bG) \geq 0] = \probaOf{ \sum_{i=1}^n \lambda_i \bG_i^2  + \mu_i \bG_i \leq \theta }.
\end{equation}
\end{framed}
Since both (\ref{eq:problem:counting}) and \Cref{theo:lishi} deal with the probability that a sum of independent random variables exceeds a threshold, the relevance of \Cref{theo:lishi} should now be clear. 

In order to prove \Cref{thm:PTF-counting} using \Cref{theo:lishi} using the formulation (\ref{eq:problem:counting}), we will need to do two preprocessing steps. These steps will be useful in~\cref{sec:upperbounds:2ptf:gaussian:sampling} as well.

\paragraph*{Pre-Processing Steps:  Discretization and Rounding:} Let us set $\gamma \eqdef 1/2^{n^{C_1}}$ and $\tau \eqdef 1/2^{n^{C_1}}$ for suitably big constants $C_1,C_2>0$.
First, without loss of generality, we may assume that the coefficients in~\eqref{eq:problem:counting} satisfy $\sum_{i=1}^n (\lambda_i^2 + \mu_i^2)=1$. We can now 
 round the coefficients $\{\lambda_i\}_{i=1}^n$ and $\{\mu_i\}_{i=1}^n$ to the nearest integral multiple of $\gamma$; we call the resulting rounded coefficients $\lambda'_i$ and $\mu'_i$. We observe that $\sum_{i=1}^n ( (\lambda_i - \lambda'_i)^2 + (\mu_i - \mu'_i)^2 ) \le O(n \gamma^2)$, and that
we also have
 \begin{equation}~\label{eq:perturb-coeff-bound}
 \frac12 \le \sum_{i=1}^n ({\lambda'}_i^{2} + {\mu'}_i^{2}) \le \frac32. 
 \end{equation}
 We now recall the following useful fact from~\cite{de2017new}:
 \begin{lemma}[{\cite[Lemma 6]{de2017new}}]~\label{lem:perturb-poly}
 Let $a(x)$ and $b(x)$ be multivariate degree-$d$ polynomials over $\mathbb{R}^n$ such that for $\mathbf{G} \leftarrow \calN$, 
 $\expect{ a(\bG) - b(\bG) }=0$, $\Var[a]=1$ 
 and $\Var[a-b] \le (\tau/d)^{3d}$. Then, 
\[
 \Prx_{\bG \sim {\cal N}} [\sign(a(\bG)) \not = \sign(a(\bG))] \le \tau. 
\]
 \end{lemma} 

Recalling that $p(x) = \sum_{i=1}^n (\lambda_i x_i^2 + \mu_i x_i) - \theta$, we define $q(x) = \sum_{i=1}^n (\lambda'_i x_i^2 + \mu'_i x_i) - \theta$. Applying~\cref{lem:perturb-poly} to $p(x)$ and $q(x)$ with $d=2$
 and $\tau = O(n^{\frac16} \cdot \gamma^{\frac13})$, we get that 
 \[
 \probaOf{ \sign\big(\sum_{i=1}^n (\lambda_i \bG_i^2  + \mu_i \bG_i) -\theta \big) \not = \sign \big( \sum_{i=1}^n (\lambda'_i \bG_i^2  + \mu'_i \bG_i) -\theta \big) } = O(n^{\frac16} \cdot \gamma^{\frac13}). 
 \]
For a suitable choice of the constant $C_1$, the right hand side is at most $O(2^{-n^2})$. Since $\Pr[\sign(p(\bG)) =1]\ge 2^{-O(n)}$,
this implies the following claim:

\begin{claim}~\label{clm:approximate-rounding}
Let $g  = \sign(p)$ and $h = \sign(q)$. For any $\eps \ge 2^{-O(n)}$, $\Pr_{\bG \leftarrow \calN} [g(\bG)=1]$ is an $(1+\eps)$-approximation of $\Pr_{\bG \leftarrow \calN} [h(\bG)=1]$. Similarly, 
$\Vert \bG_{h} -\bG_{g} \Vert_1 \le 2^{-O(n)}$. Thus,
to sample from $\bG_g$, it suffices to design an efficient sampling algorithm for $\bG_h$. 
\end{claim}

 We next define a discretization  $[\bG]_\tau := ([\bG_1]_{\tau},\dots, [\bG_n]_{\tau})$ of the Gaussian random variable $\bG = (\bG_1, \ldots, \bG_n)$ as follows: each $[\bG_i]_{\tau}$ is only supported on the points $\mathsf{AP}_{-n,n,\tau} := \{-n, -n+\tau, \ldots, n-\tau, n\}$. For each point $\kappa \in \mathsf{AP}_{-n,n,\tau}$, 
\begin{equation}~\label{eq:def-H}
[\bG_i]_{\tau}(\kappa) = \begin{cases} \Pr[\bG_i \in [\kappa, \kappa+\tau)] & \ \textrm{if} \ \kappa \not \in \{-n,n\}, \\ 
\Pr[\bG_i \in [\kappa, \infty)] & \ \textrm{if} \  \kappa = n, \\ 
\Pr[\bG_i \in (-\infty, \kappa+\tau) & \ \textrm{if}\ \kappa = -n. \\
\end{cases}
\end{equation}
In other words, we assign each point $\kappa \in \mathsf{AP}_{-n,n,\tau}$, the probability mass that the standard 
Gaussian puts in the interval $[\kappa, \kappa + \tau)$ (with the exception of $-n$ and $n$, whose probability is slightly increased by the capping).

To proceed further, for conciseness let us adopt the shorthand $\bH= (\bH_1, \ldots, \bH_n)$ to denote the random variable 
$[\bG]_\tau = ([\bG_1]_{\tau},\dots, [\bG_n]_{\tau})$. 
We next have the following claim. 
\begin{claim}~\label{clm:rounding}
For $q(x)$ as specified above, we have that
\[
\Pr_{\bG} [\sign(q(\bG)) \not = \sign(q(\bH))]  \le 2^{-\Theta(n^2)}. 
\]
\end{claim}

\begin{proof}
First of all, note that $\Pr[|\bG_i|> n] \le 2^{-\Theta(n^2)}$ for each $i \in [n]$. Let $\bE$ be the event that for all $1 \le i \le n$,  $|\bG_i| \le n$. By a union bound, we obtain
\begin{equation}~\label{eq:PrE}
 \probaOf{\overline{\bE}} = \Pr[ |\bG_i| > n \text{~for some~}i\in[n]] \le 2^{-\Theta(n^2)}. 
\end{equation}
 Using the upper bound on coefficients of $q$ from (\ref{eq:perturb-coeff-bound}), we have that 
 conditioned on event $\bE$, 
$
|q(\bG) - q(\bH)| \le O(n^2 \tau)$. Consequently, conditioned on event $\bE$, $\sign(q(\bG)) \not = \sign(q(\bH)) $ only if $|q(\bG)| = O(n^2 \cdot \tau)$. However, by anti-concentration of quadratic polynomials (\cref{thm:CW}), we have 
\begin{equation}~\label{eq:PrE2}
\Prx_{\bG \sim {\cal N}}[|q(\bG)| = O(n^2 \tau)] \le O(n \sqrt{\tau}) = 2^{-\Theta(n^2)}. 
\end{equation}
The above  application of~\cref{thm:CW} uses the lower bound on the sum of squares of the coefficients of $q$ from~\eqref{eq:perturb-coeff-bound}. Combining (\ref{eq:PrE2}) and (\ref{eq:PrE}) yields the claim. 
\end{proof}

In light of~\cref{clm:rounding,clm:approximate-rounding}, it suffices to give an algorithm to compute $\Pr_{\bH} [\sign(q(\bH))=1]$ where $q$ and $\bH$ are as defined above; we do this below.

\paragraph{Approximate counting of $\Pr_{\bH} [\sign(q(\bH))=1]$.}~\label{para:1}
 Recall that $q(\bH) = \sum_{i=1}^n ({\lambda'}_i^{2} \bH_i^2 + {\mu'}_i \bH_i) - \theta$ and that our plan is to leverage the algorithm of \cite{LS:14} (\cref{theo:lishi}) to approximate $\Pr_{\bH} [\sign(q(\bH))=1]$. Our argument employs the following three claims. 
\begin{claim}~\label{clm:discrete-LS1}
Define the random variable $\bX_i = \lambda'_i \bH_i^2 + \mu'_i \bH_i$. This random variable is supported on integral multiples of $\gamma\tau^2 = 2^{-n^{C_1} -2n^{C_2}}$ bounded in the range $[-2n^2, 2n^2]$. 
\end{claim}
\begin{proof}
The range of this random variable can be bounded just by observing that each $\lambda'_i$ and $\mu'_i$ is bounded in absolute value by $1$. Further, since $\lambda'_i$ and $\mu'_i$ are integral multiples of $\gamma$ and $\bH$ is supported on integral multiples of $\tau$, the random variable $\bX_i$ is supported on integral multiples of $\gamma \tau^2$. 
\end{proof}

\begin{claim}~\label{clm:discrete-2}
Let $\alpha$ be such that  $\Pr[\bX_i=\alpha]>0$. Then, 
$ \Pr[\bX_i = \alpha] \ge 2^{-\Theta(n^{C_1})}$. 
\end{claim}

 \begin{proof}
Note that any draw of $\bX_i$ is completely determined by the corresponding draw of $\bH_i$. Thus, it suffices to show that 
\[
\min_{\alpha: \Pr[\bH_i=\alpha]>0} \Pr[\bH_i = \alpha] \ge 2^{-\Theta(n^{C_1})}.
\]
Note that at any point $x \in [-n,n]$, the density of $\bG_i$ is at least $2^{-\Theta(n^2)}$. Note further that any support point of $\bH_i$ lies in the set $\mathsf{AP}_{-n,n,\tau}$. For any point $\alpha \neq n$ in this set, one can easily lower bound $\bH_i(\alpha)$ as 
\[
\bH_i(\alpha) = \int_{x \in [\alpha, \alpha + \tau)} \bG_i(x) dx  \geq 2^{-\Theta(n^2)} \cdot \tau \ge 2^{-\Theta(n^{C_1})}.  
\]
For $\alpha =n$, 
\[
\bH_i(\alpha) = \int_{x \ge n} \bG_i(x) dx = \int_{x \geq n} \frac{1}{\sqrt{2\pi}} e^{-\frac{x^2}{2}} dx \ge 2^{-\Theta(n^2)}. 
\]
So, for every possibility for $\alpha$, we get the claimed lower bound on $\bH_i(\alpha)$, finishing the proof. 
\end{proof}

\begin{claim}\label{claim:implementing:oracle}
Let $a, b \in \R$ where $|a|, |b| \le 1$. Let $\bY$ denote the random variable defined by $a{[\bG]_{\tau}}^2 + {b}[\bG]_{\tau}$ where $\bG$ is a standard normal random variable. 
Assuming $a$ and $b$ are rational multiples of $2^{-\poly(n)}$, we can implement $\mathcal{O}_{\bY}$ in $\poly(n)$ time. 
\end{claim}

\begin{proof}
First, we observe that given any $a$ and $b$ as above, the set 
\[
\mathcal{C}_{a,b} := \{x: \nu_1 \le ax^2 + bx  \le \nu_2\}
\]
is  given by a union of at most two closed intervals $I_1$ and $I_2$ where the boundary points of $I_1$ and $I_2$ can be (efficiently) obtained by solving quadratic equations in one-variable. Moreover, in view of the definition of $[\bG]_{\tau}$ from $\bG$, for any interval $I$ we have that $\Pr[[\bG]_\tau \in I] = \Pr[\bG \in J]$, where $J\supseteq I$ is an efficiently computable interval (depending only on the endpoints of $I$ and on $\tau$).  Finally, given any interval $J$, the value $\Pr[\bG \in J]$ can clearly be computed in polynomial time. This finishes the claim. 
\end{proof}

Given the above three claims, we are left in a position where we can apply~\cref{theo:lishi} to the discrete random variables
\[
  \bX'_i \eqdef \frac{1}{\gamma\tau^2}(\bX_i - m_i)\,, \qquad  1\leq i\leq n
\] 
where $m_i= \min_{-n\leq j \leq n} (\lambda'_i j^2 + \mu'_i j)$ (this ensures that indeed $0$ belongs to the support of $\bX'_i$). By~\cref{clm:discrete-LS1}, the $\bX'_i$'s are thus non-negative integer-valued random variables with $0\in \operatorname{supp}(\bX'_i) \subseteq \{0,1,\dots,\frac{|m_i|+2n^2}{\gamma\tau^2}\} \subseteq \{0,1,\dots,\frac{4n^2}{\gamma\tau^2}\}$.
The  parameter $C$ in \Cref{theo:lishi} is therefore upper bounded by $\frac{4n^2}{\gamma\tau^2} = 2^{n^{C_1}+2n^{C_2}+o(n)}$, so $\log C$ is indeed $\poly(n)$. Moreover, we can implement the required oracles $\{\mathcal{O}_{\bX'_i}\}_{i=1}^n$ with a $\poly(n)$-time overhead by~\cref{claim:implementing:oracle}.

 To conclude and be able to claim a $\poly(n, 1/\eps)$ runtime overall after invoking~\cref{theo:lishi}, we also require a $2^{-\poly(n)}$ bound on the parameter $\Delta = \prod_{i=1}^n \probaOf{\bX'_i = 0}$.  This bound follows from~\cref{clm:discrete-2}: for any $\alpha \in \operatorname{supp}(\bX'_1)\times\dots\times\operatorname{supp}(\bX'_n)$, we have $\prod_{i=1}^n \probaOf{\bX'_i = \alpha_i} \geq \left(2^{-\Theta(n^{C_1})} \right)^n = 2^{-\Theta(n^{C_1+1})}$. Since $0 \in \operatorname{supp}(\bX'_i)$ for all $i$, this leads to the same lower bound on $\Delta$, and applying~\cref{theo:lishi}  concludes the proof of~\cref{thm:PTF-counting}.

\subsection{Weak approximate generation: Gaussian sampling under quadratic constraints}\label{sec:upperbounds:2ptf:gaussian:sampling}

In this subsection, we will prove the following theorem.

\begin{theorem}~\label{thm:ub-sampling-ptf}
There is an efficient algorithm $\mathcal{RS}$ with the following guarantee: Given as input a degree-2 polynomial $p$ defining a PTF $g =\sign(p)$ over $\mathbb{R}^n$ such that $g^{-1}(1)$ has mass at least $2^{-n}$ under $\mathcal{N}$, and input $\eps\in(0,1]$ such that $\eps \geq 1/2^{O(n)}$, the algorithm runs in time $\poly(n/\eps)$ and outputs a point $x$ distributed according to a distribution $\mathcal{D}$ such that $\Vert \mathcal{D} - \bG_g \Vert_1 \le \eps$.
\end{theorem}

Similar to Section~\ref{sec:upperbounds:2ptf:gaussian:counting}, we can assume that $p$ is of the form 
$$
p(x) = \sum_{i=1}^n \lambda_i x_i^2 + \mu_i x_i -\theta. 
$$
As in the previous section, we set parameters $\gamma \eqdef 1/2^{n^{C_1}}$ and $\tau \eqdef 1/2^{n^{C_1}}$ for suitably big constants $C_1,C_2>0$, we define the coefficients $\lambda'_i$ and $\mu'_i$ obtained by rounding the parameters $\lambda_i$ and $\mu_i$ to the nearest integral multiple of $\gamma$ (identical to Section~\ref{sec:upperbounds:2ptf:gaussian:counting}), and we define $q(x) = 
\sum_{i=1}^n \lambda'_i x_i^2 + \sum_{i=1}^n \mu'_i x_i - \theta$ and  $h = \sign(q)$. Using Claim~\ref{clm:approximate-rounding}, it suffices to give an efficient sampling algorithm for the distribution $\bG_h$.

We define the random variable $\bH$ as in (\ref{eq:def-H}). We also define 
$\bG_{h(\bH)}$ as the random variable $\bG$ conditioned on $h(\bH)=1$. Finally, $\bH_{h}$ denotes the distribution $\bH$ conditioned on $h(\bH)=1$.

By Claim~\ref{clm:rounding}, we have that $\Vert \bG_h - \bG_{h(\bH)}\Vert_1 \le 2^{-\Theta(n^2)}$. Thus, it suffices to produce a sampler for the distribution $ \bG_{h(\bH)}$. Next, observe that since our algorithms are efficient, it suffices to sample points $ \bG_{h(\bH)}$ up to $n^{\Theta(1)}$ bits of precision.   We further recall that $\Pr[\Vert \bG \Vert_\infty > n ] \le 2^{-\Theta(n^2)}$.  Together, these observations imply that it suffices to sample from $\bH_{h}$. To accomplish this, 
we use the generic idea of reducing sampling to approximate counting~\cite{JVV86}. We begin with the following definition.

\begin{definition}~\label{def:restricted-H}
Let $\okappa^{(1)}, \ \okappa^{(2)} \in \mathsf{AP}_{-n,n,\tau}^n$ where $\okappa^{(1)}_j \le \okappa^{(2)}_j$ for every $1 \leq j \leq n$ (this is denoted by $\okappa^{(1)} \preceq \okappa^{(2)}$).
Then the random variable $\bH_{\okappa^{(1)},\okappa^{(2)}}$ represents 
$\bH$ conditioned on the $j^{th}$ coordinate lying in the interval $[\okappa^{(1)}_j , \okappa^{(2)}_j]$ for every $1 \leq j \leq n.$ 
\end{definition}

We now claim that the algorithm from the preceding subsection can be augmented to achieve the following guarantee:

\begin{theorem}~\label{label:modified-counting}
There exists a deterministic algorithm $\mathcal{AC}'$ with the following guarantee: Given a degree-$2$ polynomial $q: \mathbb{R}^n \rightarrow \mathbb{R}$ (satisfying the above conditions), error parameter $\eps \ge 2^{-O(n)}$, and vectors $\okappa^{(1)} \preceq \ \okappa^{(2)} \in \mathsf{AP}_{-n,n,\tau}^n$, algorithm $\mathcal{AC}'$ returns a $(1 \pm \eps)$ approximation to the quantity $\Pr[h(\bH_{\okappa^{(1)},\okappa^{(2)}})=1]$. 
\end{theorem}

Briefly, the algorithm in the preceding subsection relies on three crucial claims: 

\begin{enumerate}

\item The random variable $\bX_i = \lambda'_i \bH_i^2  + \mu'_i \bH_i$ is an integral multiple of $\gamma \tau^2 = 2^{-n^{C_1} -2n^{C_2}}$ bounded in the range $[-2n^2, 2n^2]$ (Claim~\ref{clm:discrete-LS1}). Here $\bH_i$ is the $i^{th}$ coordinate of $\bH$.  Let us now define $\bX_{i,\okappa^{(1)},\okappa^{(2)}}$  by replacing $\bH$ with $\bH_{\okappa^{(1)},\okappa^{(2)}}$ in the definition of $\bX_i$. 
Note that  the support of $\bH_{\okappa^{(1)},\okappa^{(2)}}$ is a subset of the support of $\bH$.  Thus, the support of $\bX_{i,\okappa^{(1)},\okappa^{(2)}}$ is a subset of the support of $\bX_i$, and as a result, Claim~\ref{clm:discrete-LS1} will continue to hold true for $\bX_{i,\okappa^{(1)},\okappa^{(2)}}$.

\item  For any $\alpha$ such that $\Pr[\bX_i = \alpha]>0$, we have that $\Pr[\bX_i=\alpha] \ge 2^{-\Theta(n^{C_1})}$ (Claim~\ref{clm:discrete-2}). It is easy to see that the same proof also implies Claim~\ref{clm:discrete-2} when $\bX_i$ is replaced by $\bX_{i,\okappa^{(1)},\okappa^{(2)}}$. 

\item Let $a,b$ be rational multiples of $2^{-\poly(n)}$ of magnitude at most 1. Let $\bY$ denote the random variable $a \bH_i^2 + b \bH_i$. Then, we can implement the oracle $\mathcal{O}_{\bY}$ in polynomial time (Claim~\ref{claim:implementing:oracle}). It is easy to see that this continues to hold if we replace $\bH_i$  by $\bH_{i,\okappa^{(1)},\okappa^{(2)}}$. 

\end{enumerate}

As the analogues of  Claim~\ref{clm:discrete-2}, Claim~\ref{clm:discrete-LS1} and Claim~\ref{claim:implementing:oracle}
hold for $\bX_{i,\okappa^{(1)},\okappa^{(2)}}$, the proof of correctness of the algorithm in the previous subsection can be modified \emph{mutatis mutandis} to 
 prove Theorem~\ref{label:modified-counting}. 

We now complete the proof of Theorem~\ref{thm:ub-sampling-ptf} by presenting and analyzing the algorithm $\mathcal{RS}.$ The algorithm is a recursive routine and is given below; it is a straightforward translation of the usual counting-to-sampling reduction to our setting (augmented with binary search since in our context we are dealing with larger domains for each coordinate than $\zo$).

\begin{enumerate}

\item[(1)] Initialize $\okappa^{\ell} = (-n,-n, \ldots, -n)$ and $\okappa^{u} = (n,n, \ldots, n)$.  

\item[(2)] Initialize  $\mathcal{A}$ as $\mathcal{A}_{\eps}= [\okappa^{\ell}_1, \okappa^{u}_1] \times \ldots \times [\okappa^{\ell}_n, \okappa^{u}_n] \cap (\mathsf{AP}_{-n,n,\tau})^n$. 

\item[(3)] Set $\delta = \eps /(2 \log |\mathcal{A}|)$.

\item[(4)] If $|\mathcal{A}|=1$, then return $\mathcal{A}$. 

\item[(5)] Else,  choose the first $j$ such that $[\okappa^{\ell}_j, 
\okappa^{u}_j] \cap \mathsf{AP}_{-n,n,\tau} >1$. Let $\eta^{\textsf{mid}} = (\okappa^{j}_{\ell} + \okappa^{j}_{u})/2$. 
Define $\okappa^{1,\textsf{mid}}$ and $\okappa^{2,\textsf{mid}}$ as
\[
\okappa^{1,\textsf{mid}} = (\okappa^{u}_1, \ldots, \okappa^{u}_{j-1}, \eta^{\textsf{mid}}, \okappa^{u}_{j+1}, \okappa^{u}_n). 
\]
\[
\okappa^{2,\textsf{mid}} = (\okappa^{\ell}_1, \ldots, \okappa^{\ell}_{j-1}, \eta^{\textsf{mid}}, \okappa^{\ell}_{j+1}, \okappa^{\ell}_n). 
\]

\item[(6)] Invoke Theorem~\ref{label:modified-counting} to obtain $(1 \pm \delta)$ approximations to the quantities  $\Pr[h(\bH_{\okappa^{1}, \okappa^{1,\textsf{mid}}})=1]$ and $\Pr[h(\bH_{\okappa^{2,\textsf{mid}}, \okappa^{2}})=1]$. 

\item[(7)] Call these approximations $\eta_1$ and $\eta_2$. Let $\mathbf{b}$ be a Bernoulli random variable where $\Pr[\mathbf{b}=0] = \eta_1 /(\eta_1 + \eta_2)$. 

\item[(8)] Define $\mathcal{A}_0$ and $\mathcal{A}_1$ as 
\[
\mathcal{A}_0 = [\okappa^{\ell}_1, \okappa^{u}_1] \times \ldots [\okappa^{\ell}_{j-1}, \okappa^{u}_{j-1}] \times [\okappa^{\ell}_{j}, \eta^{\textsf{mid}}] \times [\okappa^{\ell}_{j+1}, \okappa^{u}_{j+1}] \times \ldots [\okappa^{\ell}_n, \okappa^{u}_n];
\] 
\[
\mathcal{A}_1 = [\okappa^{\ell}_1, \okappa^{u}_1] \times \ldots [\okappa^{\ell}_{j-1}, \okappa^{u}_{j-1}] \times [\eta^{\textsf{mid}}, \okappa^{u}_{j}] \times [\okappa^{\ell}_{j+1}, \okappa^{u}_{j+1}] \times \ldots [\okappa^{\ell}_n, \okappa^{u}_n]. 
\] 

\item[(9)] Sample $b \leftarrow \mathbf{b}$ and go to Step~2 with $\mathcal{A} \leftarrow \mathcal{A}_b$.

\end{enumerate}
We now make the following claim.

\begin{claim}~\label{clm:terminate-proc1}
The above procedure terminates after $\log |\mathcal{A}|$ repetitions. The total running time is $\poly(n,1/\eps)$. 
\end{claim}
 \begin{proof}
 Note that at each stage $\mathcal{A}$ shrinks in size by a factor of $2$ and when $|\mathcal{A}|=1$, then the procedure terminates. This proves the first part of the claim. The second part follows from the fact that in each iteration every step (including Step~6) runs in time $\poly(n,1/\eps)$. 
 \end{proof}
 
Next, we observe that corresponding to any string  $z$ of length at most $m = \lceil \log|\mathcal{A}| \rceil$, we can associate a set $\mathcal{A}_z$ as follows: 
\begin{enumerate}
\item Set $\mathcal{A} \leftarrow \mathcal{A}_{\eps}$ (recall step (2) of algorithm $\mathcal{RS}$).
\item Set $j \leftarrow 1$. 
\item If $|\mathcal{A}|=1$ or $j=|z|+1$, return $\mathcal{A}_z := \mathcal{A}$.
\item Else, we define $\mathcal{A}_0$ and $\mathcal{A}_1$ by Steps (5) and (8) of the above algorithm. 
\item Let $b$ be the $j^{th}$ bit of {$z$}. 
Then, set $\mathcal{A} \leftarrow \mathcal{A}_{b}$ and $j \leftarrow j+1$. Go to Step~3. 
\end{enumerate}

We observe that (1) the above routine  associates a unique string of length at most $m$ with any element $x \in \mathcal{A}_{\eps}$. (2) If a string $z$ is a prefix of a string $w$, then $\mathcal{A}_w \subseteq \mathcal{A}_z$. 

We are now ready for the following claim:

\begin{claim}
Let $z$ be any string and let $\mathcal{A}_z \subseteq\mathcal{A}_{\eps}$ defined above. Define  $p_{\mathsf{sample},z}$ to be the probability that $\mathcal{RS}$ samples an element from $\mathcal{A}_z$. Let $p_{\mathsf{true},z} = \Pr_{\bH} [\bH \in \mathcal{A}_z| h(\bH)=1].$ Then 
\[
1- 2\delta \cdot |z| \le  \frac{p_{\mathsf{sample},z}}{p_{\mathsf{true},z}} \le 1+ 2\delta \cdot |z|. 
\]
\end{claim}

\begin{proof}
The proof is a simple induction on the length of $z$. First, observe that the bounds trivially hold when $z = \eps$. Now, inductively assume that the bounds hold when the length of $z$ is at most $\ell$. Let us consider any $z = z' \circ b$ where $b \in \{0,1\}$.
For steps (5), (6) and (7) of the above algorithm, define $\eta_{1,\mathsf{true}}$ to be $\Pr[h(\bH_{\okappa^{1}, \okappa^{1,\textsf{mid}}})=1]$ and $\eta_{2,\mathsf{true}}$ to be $\Pr[h(\bH_{\okappa^{2,\textsf{mid}}, \okappa^{2}})=1]$. Further, let $\eta_1$ and $\eta_2$ be the $(1 \pm \delta)$ approximations to $\eta_1$ and $\eta_2$ respectively that are obtained in step (6).
We have that
\[
p_{\mathsf{sample},z} = p_{\mathsf{sample},z'} \cdot \frac{\eta_{b+1}}{\eta_1 + \eta_2}  \quad \text{and} \quad p_{\mathsf{true},z} = p_{\mathsf{true},z'}  \cdot \frac{\eta_{b+1,\mathsf{true}}}{\eta_{1,\mathsf{true}} + \eta_{2,\mathsf{true}}}. 
\]
Using the induction hypothesis  for $z'$ and the fact that $\eta_b$ is a $(1 \pm \delta)$ approximation of $\eta_{b,\mathsf{true}}$, we get the claim for $z$. 
\end{proof}

To complete the proof of \Cref{thm:ub-sampling-ptf}, it remains only to recall that $\delta = \eps/2m$.

\section{Hardness results for learning from satisfying assignments}\label{sec:lowerbounds}

In this section we show that under various standard cryptographic assumptions, our algorithmic results from the previous sections are close to the strongest possible for learning with respect to log-concave and normal background distributions.  In particular, we will show that under cryptographic assumptions, there is no efficient algorithm to 

\begin{enumerate}

\item learn the class of $n$-variable degree-two polynomial threshold functions $\mathsf{PTF}_2^n$ when the background distribution is a known log-concave distribution. In fact, this lower bound holds even when the background distribution is the uniform distribution over the solid $n$-dimensional hypercube $[0,1]^n$ (which is an extremely simple log-concave distribution).

\item learn the class of $n$-variable degree-four polynomial threshold functions $\mathsf{PTF}_4^n$ when the background distribution is the $n$-dimensional standard Gaussian $\calN(0,1)^n.$

\end{enumerate}

We obtain these results using (an extension of) a general condition for showing hardness of learning from~\cite{DDS:15}. We begin by recalling the notion of an \emph{invertible Levin reduction}.

\begin{definition}
A binary relation $R$ is said to reduce to another binary relation $R_0$ by a polynomial-time invertible
Levin reduction if there are three algorithms $\alpha(\cdot)$, $\beta(\cdot,\cdot)$ and $\gamma(\cdot,\cdot)$, each running in polynomial time, with the following properties:

\begin{enumerate}

\item For each $x,y$, if $(x,y) \in R$ then $(\alpha(x), \beta(x,y)) \in R_0$. 

\item For each $x,z$, if $(\alpha (x),z) \in R_0$ then $(x, \gamma(\alpha(x),z)) \in R.$

\item Finally, the functions $\beta$ and $\gamma$ are such that for each $x,y$, we have $\gamma(\alpha(x),\beta(x,y))=y$.

\end{enumerate}

\end{definition}

In the context of this paper, we note that for any class $\class$ of Boolean functions (over any domain), we can define a relation $R_{\class}$ which contains those pairs $(f,z)$ such that $f \in \class$ and $f(z)=1$.  In this section, whenever we say that there is an invertible Levin reduction from class $\class_1$ to class $\class_2$, we mean that there is an invertible Levin reduction from $\mathcal{R}_{\class_1}$ to $\mathcal{R}_{\class_2}$. 
As an illustrative example, we may take $\calC_1$ to be the class of all polynomial-size Boolean circuits (corresponding to the \textsf{Circuit-SAT} problem) and $\calC_2$ to be the class of all 3-CNF formulas.
For property (1), given a circuit $C$ and a satisfying assignment $z$, the output $\alpha(C)$ is the 3-CNF formula (over an expanded variable space) that is produced by the standard reduction, and if $(C,z)$ belongs to $\mathcal{R}_{\calC_1}$ (meaning that $z$ is a satisfying assignment for circuit $C$) then the output $\beta(C,z)$ is the corresponding satisfying assignment for the 3-CNF formula $\alpha(C).$ For property (2), given a 3-CNF $\alpha(C)$ and a satisfying assignment $z'$ of that 3CNF, $\gamma(\alpha(C),z')$ outputs the (unique) satisfying assignment $z$ of $C$ that ``gave rise to'' $z'$, and it is easy to see that property (3) holds.

Next, we recall a standard assumption from the cryptographic literature~\cite{MRV99}, which is a slight variant of the standard RSA assumption. 

\begin{assumption}\label{assumption:crypto:hardness}
Let $\mathsf{RSA}_k$ be the set of all integers which are products of two primes of length $\lfloor (k-1) /2 \rfloor$. Let $\bmsg$ be chosen uniformly from $\mathsf{RSA}_k$ and let $\bx$ be chosen uniformly from $\mathbb{Z}_{\bmsg}^\ast$. Let $\bp$ be a uniformly chosen prime of length $k+1$. There exists some absolute constant $\delta>0$ such that, for any probabilistic algorithm $A$ running in time $2^{n^{\delta}}$,
\[
\Pr_{\bmsg,\bx,\bp}[A(\bmsg,\bx,\bp) = y \ \text{and} \ y^p = \bx \!\mod \bmsg]   \le 2^{-n^{\delta}}. 
\]
\end{assumption}
Under the above assumption~\cite{MRV99} showed the existence of a so-called unique signature scheme which is secure for subexponential time algorithms under ``random message attack'' (RMA).
In~\cite{DDS:15}, the authors showed how the existence of such unique signature schemes implies hardness of learning under the uniform background distribution over $\bn$ for classes $\class$ for which there is an invertible Levin reduction from $\textsf{Circuit-SAT}$. More precisely,~\cite{DDS:15} established the following theorem. 

\begin{theorem} \label{thm:DDShard}
Let $\class$ be a class of functions such that there is a polynomial-time invertible Levin reduction from \textsf{Circuit-SAT} to $\class$. Then, under~\cref{assumption:crypto:hardness}, there exists an absolute constant $\delta'>0$ such that no $2^{n^{\delta'}}$-time algorithm can learn $\class$ under the uniform background distribution. 
\end{theorem}

In the current work we will need a slight extension of~\cref{thm:DDShard} which is suitable for establishing hardness when the background distribution is something other than the uniform distribution on $\bn$. The precise result we shall use is stated below.

\begin{theorem} \label{thm:our-hard}
Let $\class$ be a class of functions from $\R^n$ to $\bits$ such that there is a polynomial-time invertible Levin reduction from \textsf{Circuit-SAT} to $\class$.  Suppose further that the background distribution $\calD$ and the reduction are such that the following properties hold: 

\begin{enumerate}
  \item\label{theo:new:hardness:item1} There is an efficient algorithm which, on input a point $x$, outputs the value of the probability density function for ${\cal D}$ at $x$ (i.e.,~an efficient algorithm to simulate an evaluation oracle for ${\cal D}$).
  \item\label{theo:new:hardness:item2}
  If $C$ is a circuit which is an instance of \textsf{Circuit-SAT} and $f \in \class$ is an instance of ${\cal C}$ which arises from $C$ in the reduction, then $f^{-1}(1)$ 
  is the disjoint union of some collection of $K$ sets $S_1,\dots,S_K$, where $K$ is the number of satisfying assignments for $C$ and each set $S_i$ corresponds to precisely one solution to $C$.  Moreover,
    \begin{enumerate}
      \item For any $i,j\leq K$, the regions $S_i$ and $S_j$ have probability mass within a factor of two of each other under ${\cal D}$; and
      \item There is an efficient algorithm which, for any $j \in [K]$, given any $x \in S_j$, outputs a sample drawn from ${\cal D}_{S_j}$ (the distribution ${\cal D}$ restricted to $S_j$).
    \end{enumerate}
\end{enumerate}

\noindent
Then, under~\cref{assumption:crypto:hardness}, there exists an absolute constant $\delta'>0$ such that no $2^{n^{\delta'}}$-time algorithm can learn $\class$ under the background distribution $\calD$. 

\end{theorem}

Inspection of the proof of~\cref{thm:DDShard} shows that the arguments used in its proof straightforwardly extend to yield~\cref{thm:our-hard}.  Intuitively,~\cref{theo:new:hardness:item1} ensures that given the hypothesis distribution generated by a learning algorithm for $\class$ with respect to background distribution ${\cal D}$ (recall that in our model such a hypothesis distribution requires query access to an evaluation oracle for ${\cal D}$), it is indeed possible to generate samples from a distribution that is statistically close to ${\cal D}_f$.  Intuitively,~\cref{theo:new:hardness:item2}, together with the invertible Levin reduction, makes it possible to translate a draw from the hypothesis distribution back to a signed message and thereby contradict~\cref{assumption:crypto:hardness}.

\subsection{A lower bound for degree-2 PTFs under log-concave background distributions}\label{sec:lowerbounds:2ptf:logconcave}

In this subsection we prove~\cref{thm:neg-deg2}. 

\begin{theorem} [\cref{thm:neg-deg2}, restated] \label{thm:neg-deg2-2}
Under~\cref{assumption:crypto:hardness},
there is no subexponential-time algorithm $A$ for learning the class $\ptf{n}{2}$ of $n$-variable degree-two polynomial threshold functions with respect to the (log-concave) background distribution $\calD$ which is uniform over the solid cube $[0,1]^n$.
\end{theorem}

We prove~\cref{thm:neg-deg2-2} using~\cref{thm:our-hard} and a reduction from \textsf{Subset-Sum}.  Recall that an instance $W$ of the \textsf{Subset-Sum} problem is given by a non-negative integer $w_0$ and an $n$-tuple $w=(w_1,\dots,w_n)$ of non-negative integers, and the problem is to determine whether there is a subset $S \subseteq [n]$ such that $w_0 = \sum_{i \in S} w_i.$  It is well known that there is a polynomial-time invertible reduction from \textsf{Circuit-Sat} to \textsf{Subset-Sum} in which a \textsf{Circuit-Sat} instance of size $n$ is mapped to a \textsf{Subset-Sum} instance in which each $w_i$ is at most $2^{n^2}.$  Since polynomial-time invertible Levin reductions compose, to prove~\cref{thm:neg-deg2-2} it suffices for us to give a polynomial-time invertible Levin reduction from \textsf{Subset-Sum} to the class of degree-2 PTFs which satisfies the properties stated in~\cref{thm:our-hard}.

\paragraph{The reduction from \textsf{Subset-Sum}.}

Given an instance $W=(w_0,w)\in\N\times\N^{n}$ of \textsf{Subset-Sum} which has each $w_i \leq 2^{n^2}$, define the quadratic polynomial
\begin{equation}\label{eq:murty:quadratic:polynomial}
  p_W(x) = \left(\sum_{i=1}^n w_i x_i - w_0\right)^2 + \lambda \sum_{i=1}^n x_i(1-x_i)\,,
\end{equation}
where $\lambda\eqdef M\cdot \normtwo{w} \gg 1$ for some $M=M(n)=cn$ to be determined later. 
The first observation is that the points achieving the minimum of $p_W$ over $[0,1]^n$ are in one-to-one correspondence with the solutions of $W$. Specifically, we observe the following.
\begin{proposition}\label{prop:subsetsum:corresp}
Suppose that the \textsf{Subset-Sum} instance $W$ is satisfiable. Then the minimum of $p_W$ over $[0,1]^n$ is $0$, and this minimum is achieved exactly at the points $x\in\zo^n$ satisfying $\dotprod{w}{x} = w_0$.
\end{proposition}
\begin{proof}
  It is immediate to see that $p_W\geq 0$ on $[0,1]^n$, and further that for any solution of $W$ we have $p_W(x)=0$ (as the first term is zero by assumption, and the second is zero since $x\in\zo^n$). Conversely, consider any $x\in[0,1]^n$ such that $p_W(x)=0$. This implies both that $w \cdot x = w_0$ (since the first term has to be zero) and that $\sum_{i=1}^n x_i(1-x_i)=0$, which since $x\in[0,1]^n$ yields $x\in\zo^n$.
\end{proof}

Based on~\cref{prop:subsetsum:corresp}, a natural choice of degree-two PTF for our hardness reduction would be, given an instance $W$, to take $f_W(x) \eqdef \sign( -p_W(x) )$ (recall that $\sign(0)$ is defined to be 1). Indeed, in this case the satisfying assignments of $f_W$ in $[0,1]^n$ are in one-to-one correspondence with the solutions of the \textsf{Subset-Sum} instance, and given $x\in [0,1]^n\cap f_W^{-1}(1)$ it is trivial to produce the corresponding solution. We do not pursue exactly this simple approach, though, because of the issue discussed in Remark~\ref{remark:reasonable}, namely that $[0,1]^n\cap f_W^{-1}(1)$ has measure zero under the uniform distribution on $[0,1]^n$ (as it is a finite set of discrete points). Since this set has measure zero, such an instance of the degree-2 PTF learning problem violates our requirement that the target function have a $2^{-n^{O(1)}}$ fraction of satisfying assignments under the background distribution.

We get around this issue by considering a slightly modified version of the above polynomial.  More precisely, we will instead define
\begin{equation}\label{eq:subsetsum:fw}
    f_W(x) \eqdef \sign\left( \frac{1}{2} - p_W(x) \right),
\end{equation}
and we will establish the following:
\begin{proposition}\label{prop:subsetsum:densities}
  With $f_W$ defined as in~\cref{eq:subsetsum:fw}, we have that

  \begin{enumerate}
   
    \item $[0,1]^n\cap f_W^{-1}(1)$ has measure at least $1/2^{O(n^3)}$ under $\uniform_{[0,1]^n}$;
   
    \item $[0,1]^n\cap f_W^{-1}(1)$ is the disjoint union of $K$ sets $S_1,\dots,S_K$, each with volume within a factor $2$ of each other under $\uniform_{[0,1]^n}$, where $K$ is the number of solutions to the \textsf{Subset-Sum} instance $W$ (and each set $S_i$ contains precisely one solution to $W$); and
   
    \item There is an efficient algorithm which, for any $j \in [K]$, given any $x \in S_j$, outputs a uniform sample drawn from $S_j$.
  \end{enumerate}
\end{proposition} 

Part~1 of~\cref{prop:subsetsum:densities} addresses the ``reasonableness'' condition discussed in~\cref{remark:reasonable}. Part~2, together with our earlier discussion, is easily seen to give the desired polynomial-time invertible Levin reduction from \textsf{Subset-Sum} to degree-2 PTFs satisfying the properties stated in~\cref{thm:our-hard}. 

 Towards proving~\cref{prop:subsetsum:densities} (which we will do at the end of this subsection), let us rewrite~\cref{eq:murty:quadratic:polynomial} as
\begin{equation}
  p_W(x) = s_W(x) + \lambda B(x)\,
  \end{equation}
  with
  \[
  s_W(x) = \left(\sum_{i=1}^n w_i x_i - w_0\right)^2
  \text{~and~}B(x)=\sum_{i=1}^n x_i(1-x_i).
  \]

Our next three claims will establish that (i) $f_W$ outputs $-1$ on any point too far from the Boolean hypercube (\cref{claim:subsetsum:farfromall}); (ii) $f_W$ outputs $-1$ on any point close to a point of the Boolean hypercube which is \emph{not} a solution to the \textsf{Subset-Sum} instance $W$ (\cref{claim:subsetsum:closetobad}); and (iii) $f_W$ outputs $1$ on any point close to a point of the Boolean hypercube which \emph{is} a solution to the \textsf{Subset-Sum} instance $W$ (\cref{claim:subsetsum:closetogood}). Combining the three (for suitable notions of ``too close'' and ``too far'' along with corresponding volume considerations will establish the proposition.

\noindent Define the parameter $\alpha_n$ as
\begin{equation}\label{eq:subsetsum:alpha}
    \alpha_n \eqdef \frac{1}{2}\left( 1-\sqrt{1-\frac{2}{\lambda}}\right) = \frac{1}{2\lambda} + \bigO{\frac{1}{\lambda^2}}
\end{equation}
\begin{claim}\label{claim:subsetsum:farfromall}
    Suppose $x\in[0,1]^n$ is at $\lp[1]$ distance more than $\alpha_n$ from every point from $\zo^n$. Then $f_W(x) = -1$.
\end{claim}
\begin{proof}
  By symmetry, it is enough to consider the point $0_n\in \zo^n$. For any $x\in[0,1]^n$ such that $\alpha_n < \normone{x} = \normone{x-0_n} < 1/2$, we have
  \[
      B(x)  = \normone{x} - \normtwo{x}^2 \geq \normone{x} - \normone{x}^2 > \alpha_n\left(1-\alpha_n\right) = \frac{1}{2\lambda}
  \]
by our choice of $\alpha_n$ as the solution to the equation $X(1-X)={\frac 1 {2 \lambda}}$,
  and therefore $p_W(x) \geq \lambda B(x) > 1/2$, so $f_W(x)=-1.$
\end{proof}

\begin{claim}\label{claim:subsetsum:closetobad}
    Let $x\in[0,1]^n$, and suppose that $x$ is at $\lp[1]$ distance at most $1/(4\normtwo{w})$ of some $z\in\zo^n$ (thus necessarily unique). If $z$ is not a solution to the \textsf{Subset-Sum} instance $W$, then $f_W(x) = -1$.
\end{claim}
\begin{proof}
  We can write
  \begin{align*}
      1&\leq \abs{ \sum_{i=1}^n w_i z_i  - w_0 } \leq \abs{ \sum_{i=1}^n w_i z_i - \sum_{i=1}^n w_i x_i }  + \abs{ \sum_{i=1}^n w_i x_i  - w_0 }
      = \abs{ \sum_{i=1}^n w_i (z_i - x_i) }  + \abs{ \sum_{i=1}^n w_i x_i  - w_0 } \\
      &\leq \normtwo{w} \cdot \normtwo{x-z} + \abs{ \sum_{i=1}^n w_i x_i  - w_0 } \leq \normtwo{w} \cdot \normone{x-z} + \abs{ \sum_{i=1}^n w_i x_i  - w_0 }
      \leq \frac{1}{4} + \sqrt{s_W(x)}
  \end{align*}
  the first inequality from the fact that all $w_i$'s are positive integers and $z\in \zo^n$, the third-to-last from Cauchy--Schwarz, and the second-to-last from monotonicity of $\lp[p]$ norms. Therefore, we have $p_W(x) \geq s_W(x) \geq (3/4)^2 > 1/2$, and $f_W(x)=-1$.
\end{proof}
\noindent(Note that since $\alpha_n \operatorname*{\sim}\limits_{n\to\infty} \frac{1}{2M\normtwo{w}} = \littleO{1/\normtwo{w}}$,~\cref{claim:subsetsum:closetobad} obviously implies that any $x\in [0,1]$ which is at $\lp[1]$ distance at most $\alpha_n$ from a non-solution point of $\zo^n$ must have $f_W(x) = -1$.)\medskip

\noindent Define the parameter $\beta_n$ as
\begin{equation}\label{eq:subsetsum:beta}
    \beta_n \eqdef \frac{1}{2\normtwo{w}}\left( \sqrt{M^2+2} - M\right) = \frac{1}{2 M\normtwo{w}} - \bigO{\frac{1}{M^3\normtwo{w}}}
\end{equation}
\begin{claim}\label{claim:subsetsum:closetogood}
    Let $x\in[0,1]^n$, and suppose that $x$ is at $\lp[1]$ distance at most $\beta_n$ from some $z\in\zo^n$ (thus necessarily unique). If $z$ is a solution to the \textsf{Subset-Sum} instance $W$, then $f_W(x) = 1$.
\end{claim}
\begin{proof}
  We can write
  \begin{align*}
      \abs{ \sum_{i=1}^n w_i x_i  - w_0 } \leq \abs{ \sum_{i=1}^n w_i (x_i  - z_i) } + \abs{ \sum_{i=1}^n w_i z_i  - w_0 }
      = \abs{ \sum_{i=1}^n w_i (x_i  - z_i) } \leq \normtwo{w} \cdot \normone{x-z}
  \end{align*}
  the last inequality again from Cauchy--Schwarz and monotonicity of $\lp[p]$ norms. Moreover, we also have
  \[
      B(x)  = \sum_{i=1}^n x_i(1-x_i) \leq  \sum_{i=1}^n \min(x_i, 1-x_i) = \sum_{i=1}^n \abs{x_i-z_i} = \normone{x-z}
  \]
  so that
  \begin{align*}
      p_W(x) 
      &= s_W(x) + \lambda B(x) 
      \leq \normtwo{w}^2 \cdot \normone{x-z}^2 + M \normtwo{w} \cdot \normone{x-z}
      = (\normtwo{w} \cdot \normone{x-z})^2 + M (\normtwo{w} \cdot\normone{x-z}) \\
      &\leq (\beta_n \normtwo{w})^2 + M (\beta_n \normtwo{w})
      = \frac{1}{2},\,
  \end{align*}
where the last inequality is a direct consequence of our choice of $\beta_n$ such that $\beta_n \normtwo{w}$ is a solution to $X^2+MX=1/2$. This implies $f_W(x)=1$, as claimed.
\end{proof}

\begin{proofof}{\cref{prop:subsetsum:densities}}
  \cref{claim:subsetsum:farfromall,claim:subsetsum:closetobad} together imply (since, as noted after~\cref{claim:subsetsum:closetobad}, $1/(4\normtwo{w} \gg \alpha_n$) that if $x\in[0,1]^n$ is not within $\lp[1]$ distance $\alpha_n$ of a solution $z\in\zo^n$ of the \textsf{Subset-Sum} instance $W$, then $f_W(x) = -1$. Moreover,~\cref{claim:subsetsum:closetogood} ensures that any $x\in[0,1]^n$ which is within $\lp[1]$ distance $\beta_n$ of a solution $z\in\zo^n$ of the \textsf{Subset-Sum} instance $W$ satisfies $f_W(x) = 1$.
  
  Therefore, as $\alpha_n = o(1)$ and any two point in $\zo^n$ are at least at unit $\lp[1]$ distance, the satisfying assignments $f_W^{-1}(1)$ are disjoint sets $S_1,\dots, S_K$ (one around each of the $K$ satisfying assignments $z_1,\dots, z_K\in\zo^n$ of the \textsf{Subset-Sum} instance) such that, for all $1\leq k\leq K$,
  \begin{equation*}
          B^1_n( z_k, \beta_n )\cap [0,1]^n \subseteq S_k \subseteq B^1_n( z_k, \alpha_n )\cap [0,1]^n\,.
  \end{equation*}
  This gives item~(2) of the proposition, except for the assertion that any two $S_i,S_j$ have volume within a factor of two of each other (we will establish this below).
  
  For items (1) and (3), by symmetry it suffices to consider the case $z_k=0_n$. Noting that the volume of the $\lp[1]$ ball of radius $r$ is $\operatorname*{Leb}(B^1_n( 0_n, r )) = \frac{2^n r^n}{n!}$, and that restricting this ball to $[0,1]^n$ yields an $1/2^n$ fraction of the original volume, we get $\operatorname*{Leb}(B^1_n( 0_n, r )\cap[0,1]^n) = \frac{r^n}{n!}$ and therefore by~\cref{claim:subsetsum:farfromall,claim:subsetsum:closetogood} we have
  \begin{equation*}
          \frac{\beta_n^n}{n!} \leq \operatorname*{Leb}(S_k) \leq \frac{\alpha_n^n}{n!}\,.
  \end{equation*}
  From~\cref{eq:subsetsum:alpha,eq:subsetsum:beta} and our setting of $\lambda=M\normtwo{w}$, we get that $\alpha_n,\beta_n$ are respectively $\frac{1}{2\lambda}+\bigO{\frac{1}{M\lambda}}$ and $\frac{1}{2\lambda}-\bigO{\frac{1}{M\lambda}}$, from which, as long as $M = cn$ (for some sufficiently big absolute constant $c>0$ to make $c'>0$ below small enough), we get
  \begin{equation}
          \frac{1}{\sqrt{2}\cdot  2^n \lambda^n n!} \leq \left(1-\frac{c'}{n}\right)^n\frac{1}{2^n \lambda^n n!} \leq 
           \frac{\beta_n^n}{n!} \leq
           \operatorname*{Leb}(S_k) 
           \leq
           \frac{\alpha_n^n}{n!}
           \leq \left(1+\frac{c'}{n}\right)^n\frac{1}{2^n \lambda^n n!} \leq \frac{\sqrt{2}}{2^n \lambda^n n!} \,.
  \end{equation}
  This shows that all volumes $\operatorname*{Leb}(S_k)$ are within a constant factor $2$ of each other, finishing item~2 of the Proposition. Moreover, under the uniform distribution $\uniform_{[0,1]^n}$ over  $[0,1]^n$, the above along with our bound $\norminf{w}\leq 2^{n^2}$ and the choice of $\lambda=M \cdot \normtwo{w}$ implies that
  \begin{equation}
          \uniform_{[0,1]^n}(S_k) \geq \frac{1}{\sqrt{2}\cdot 2^n\lambda^n n!} = \frac{1}{2^{O(n^3)}} \,,
  \end{equation}
  as $\lambda \leq M \sqrt{n}\norminf{w} = O(n^{3/2}2^{n^2})$.
  
  This establishes item~1 of \cref{prop:subsetsum:densities}, so it remains only to show that item~3 holds. To establish this last item (i.e., that for any $k$ one can efficiently, given any $x^\ast\in S_k$, sample uniformly from $S_k$) it suffices to observe that (i) given such a $x^\ast$, it is immediate to find the corresponding satisfying assignment $z_k\in\zo^n$ (by rounding the coordinates); (ii) efficient sampling uniformly from $B^1_n( z_k, \alpha_n )$ can be done using elementary techniques; and (iii) since the volume of $S_k$ is within a factor of $2$ of the volume of $B^1_n( z_k, \alpha_n )$ and $f_W$ can be efficiently evaluated, rejection sampling allows us to sample uniformly from $S_k$ in an efficient way.
  This concludes the proof of \cref{prop:subsetsum:densities} and hence also of \cref{thm:neg-deg2-2}.
\end{proofof}
   
\subsection{A lower bound for degree-4 PTFs under the normal distribution}\label{sec:lowerbounds:4ptf:gaussian}

In this subsection, we prove~\cref{thm:neg-deg4}, which shows that even under a very strong assumption on the ambient distribution one cannot hope for much better than~\cref{thm:pos-deg2}:
\begin{theorem} [\cref{thm:neg-deg4}, restated] \label{thm:neg-deg4-2}
Under~\cref{assumption:crypto:hardness}, there is no subexponential-time algorithm $A$ for learning the class $\ptf{n}{4}$ of $n$-variable degree-four polynomial threshold functions with respect to the standard normal distribution $\mathcal{N}(0,1)^n$.
\end{theorem}
The rest of this section is dedicated to the proof of this theorem. As for~\cref{thm:neg-deg2}, our starting point is a reduction from~\textsf{Subset-Sum}, although a slightly different one. First off, it will be convenient for us to now view the \textsf{Subset-Sum} problem as being over $\{-1,1\}^n$ rather than $\{0,1\}^n$: i.e., the goal is to determine whether there is a string $z \in \{-1,1\}^n$ such that $w \cdot z = w_0$.  (It is easy to see that this is equivalent to the original problem over $\{0,1\}^n$.)  Now, given an instance $W=(w_0,w)\in\N\times\N^{n}$ of this form of \textsf{Subset-Sum} in which each $w_i \leq 2^{n^2}$, we define the quartic polynomial
\begin{equation}\label{eq:deg4:polynomial}
  p_W(x) = \left(\sum_{i=1}^n w_i x_i - w_0\right)^2 + \lambda \sum_{i=1}^n (x_i^2-1)^2 = s_W(x) + \lambda B(x)\,,
\end{equation}
where $\lambda\eqdef cn\max(\normtwo{w}^2,n) \gg 1$ for some sufficiently large absolute constant $c>0$, and we define 
\begin{equation}\label{eq:deg4:subsetsum:fw}
    f_W(x) \eqdef \sign\left( \frac{1}{2} - p_W(x) \right),
\end{equation}
as in~\cref{thm:neg-deg2}. Compared to the degree-2 lower bound, the ``penalization'' term $B(x)$ is now a degree-4 polynomial, whose role is to enforce that satisfying assignments to $f_W$ from $\R^n$ must be ``essentially Boolean'', where now ``Boolean'' means ``belonging to $\{-1,1\}^n$'' (note that $B$ vanishes exactly on $\{-1,1\}^n$) over inputs from $\R^n$).

Analogously to the proof of~\cref{thm:neg-deg2}, we establish the facts that (1)~zeros of $p_W$ are exactly the solutions to the~\textsf{Subset-Sum} instance $W$; (2)~if a point is far from a Boolean point, then it does not satisfy $f_W$; (3)~if a point is close to a Boolean point not satisfying $W$, then it does not satisfy $f_W$; and (4)~if a point is close to a Boolean point satisfying $W$, then it does satisfy $f_W$. Altogether, this will in turn allow us to establish the proposition below, which is closely analogous to~\cref{prop:subsetsum:densities}:

\begin{proposition}\label{prop:deg4:subsetsum:densities}
  With $f_W$ defined as in~\cref{eq:deg4:subsetsum:fw}, we have that
  \begin{enumerate}
    \item $f_W^{-1}(1)$ has measure at least $1/2^{O(n^3)}$ under $\mathcal{N}(0,1)^n$;
   
    \item $f_W^{-1}(1)$ is the disjoint union of $K$ sets $S_1,\dots,S_K$, each with measure within a factor $2$ of each other under $\mathcal{N}(0,1)^n$, where $K$ is the number of solutions to the \textsf{Subset-Sum} instance $W$ (and each set $S_i$ contains precisely one solution to $W$); and
   
    \item There is an efficient algorithm which, for any $j \in [K]$, given any $x \in S_j$, outputs a draw from ${\cal N}(0,1)^n$ restricted to $S_j$.
  \end{enumerate}
\end{proposition}

Similar to \cref{prop:subsetsum:densities}, \cref{prop:deg4:subsetsum:densities} is easily seen to satisfy the ``reasonableness'' condition and give the desired polynomial-time invertible Levin reduction from \textsf{Subset-Sum} to degree-2 PTFs satisfying the properties stated in~\cref{thm:our-hard}. 
The following is entirely analogous to~\cref{prop:subsetsum:corresp}:

\begin{proposition}\label{obs:deg4:subsetsum:corresp}
Suppose that the \textsf{Subset-Sum} instance $W$ is satisfiable. Then the minimum of $p_W$ over $\R^n$ is $0$, and this minimum is achieved exactly at the points $x\in\{-1,1\}^n$ satisfying $\dotprod{w}{x} = w_0$.
\end{proposition}
\begin{proof}
  For any solution $x$ of $W$ we have $p_W(x)=0$ (as the first term $s_W(x)$ is zero by assumption, and the penalization term $B(x)$ is zero since $x\in\{-1,1\}^n$). Conversely, consider any $x\in\R^n$ such that $p_W(x)=0$. Since $s_W,B\geq 0$, this implies both that $w \cdot x = w_0$ (since the first term has to be zero) and that $\sum_{i=1}^n (x_i^2-1)^2=0$, which yields $x\in\{-1,1\}^n$.
\end{proof}

As before, our actual construction works with the PTF $\sign({\frac 1 2} - p_W(x))$ rather than $\sign(-p_W(x))$ because of the requirement that the learning problems under consideration have at least an inverse exponential fraction of probability mass under the background distribution lying on satisfying assignments of the target function (recall~\cref{remark:reasonable}).
Define the parameter $\alpha_n$ as the smallest positive solution to to $4(1-X)^2 X^2 \geq 1/(2\lambda)$, i.e.,
\begin{equation}\label{eq:deg4:subsetsum:alpha}
    \alpha_n \eqdef \frac{1}{2}\left(1-\sqrt{1-\sqrt{\frac{2}{\lambda}}}\right) = \frac{1}{\sqrt{8\lambda}}+\frac{1}{8\lambda} + \littleO{\frac{1}{\lambda}}\,.
\end{equation}
\begin{claim}\label{claim:deg4:subsetsum:farfromall}
    Suppose $x\in\R^n$ is at $\lp[2]$ distance more than $\alpha_n$ from every point from $\{-1,1\}^n$. Then $f_W(x) = -1$.
\end{claim}
\begin{proof}
  For any such $x\in\R^n$, we have that the closest Boolean point to $x$ is $x^\ast\in\{-1,1\}^n$, defined by $x^\ast_i = \sign(x_i)$ for every $i\in[n]$. Then, 
  \begin{align*}
      B(x)  &= \sum_{i=1}^n (x_i-1)^2(x_i+1)^2 = \sum_{i=1}^n (x_i-x^\ast_i)^2(x_i+x^\ast_i)^2\,.
  \end{align*}
  Since $(x_i+x^\ast_i)^2\geq 1$ for all $i$, we first observe that if there exists some $i\in[n]$ such that $\abs{x_i-x^\ast_i} > 2\alpha_n$, then $B(x) > 4\alpha_n^2 > 1/(2\lambda)$, and we are done. Thus we may assume that $\abs{x_i-x^\ast_i} < 2\alpha_n$ for all $i$. It follows that
  \begin{align*}
      B(x)  &>\sum_{i=1}^n (x_i-x^\ast_i)^2(2-2\alpha_n)^2 = 4(1-\alpha_n)^2\sum_{i=1}^n (x_i-x^\ast_i)^2 > 4(1-\alpha_n)^2\alpha_n^2\,,
  \end{align*} 
 where the first inequality uses $\abs{x_i-x^\ast_i} < 2\alpha_n$ and the last inequality follows by recalling that $\normtwo{x-x^\ast}^2 > \alpha_n^2$.
  We then get $p_W(x) \geq \lambda B(x) > 1/2$, so $f_W(x)=-1.$
\end{proof}

\begin{claim}\label{claim:deg4:subsetsum:closetobad}
    Let $x\in\R^n$, and suppose that $x$ is at $\lp[2]$ distance at most $1/(4\normtwo{w})$ from some $z\in\{-1,1\}^n$ (thus necessarily unique). If $z$ is not a solution to the \textsf{Subset-Sum} instance $W$, then $f_W(x) = -1$.
\end{claim}
\begin{proof}
  The proof is almost identical to that of~\cref{claim:subsetsum:closetobad}. We write
  \begin{align*}
      1&\leq \abs{ \sum_{i=1}^n w_i z_i  - w_0 } \leq \abs{ \sum_{i=1}^n w_i z_i - \sum_{i=1}^n w_i x_i }  + \abs{ \sum_{i=1}^n w_i x_i  - w_0 }
      \leq \normtwo{w}\normtwo{x-z} + \sqrt{s_W(x)}
      \leq \frac{1}{4} + \sqrt{s_W(x)}
  \end{align*}
  the first inequality from the fact that all $w_i$'s are positive integers and $z\in \{-1,1\}^n$, the third from Cauchy--Schwarz. Therefore, we have $p_W(x) \geq s_W(x) \geq (3/4)^2 > 1/2$, and $f_W(x)=-1$.
\end{proof}

\noindent Define the parameter $\beta_n$ as the smallest positive solution to $(\normtwo{w}^2+\lambda(2+X)^2)X^2=1/2$, so that
\begin{equation}\label{eq:deg4:subsetsum:beta}
    \beta_n \eqdef \frac{1}{\sqrt{2(\normtwo{w}^2+4\lambda)}} - \bigO{\frac{1}{\lambda}}.
\end{equation}
\begin{claim}\label{claim:deg4:subsetsum:closetogood}
    Let $x\in\R^n$, and suppose that $x$ is at $\lp[2]$ distance at most $\beta_n$ from some $z\in\{-1,1\}^n$ (thus necessarily unique). If $z$ is a solution to the \textsf{Subset-Sum} instance $W$, then $f_W(x) = 1$.
\end{claim}
\begin{proof}
  We can write
  \begin{align*}
      \abs{ \sum_{i=1}^n w_i x_i  - w_0 } \leq \abs{ \sum_{i=1}^n w_i (x_i  - z_i) } + \abs{ \sum_{i=1}^n w_i z_i  - w_0 }
      = \abs{ \sum_{i=1}^n w_i (x_i  - z_i) } \leq \normtwo{w} \cdot \normtwo{x-z} \leq \normtwo{w}\beta_n,
  \end{align*}
  the second-to-last inequality again from Cauchy--Schwarz. Moreover, we also have
  \[
      B(x)  = \sum_{i=1}^n (x_i-1)^2(x_i+1)^2  = \sum_{i=1}^n (x_i-z_i)^2(x_i+z_i)^2 
      \leq  (2+\beta_n)^2\sum_{i=1}^n (x_i-z_i)^2 \leq (2+\beta_n)^2\beta_n^2
  \]
  where we used the fact that $\abs{x_i+z_i}\leq 2+\beta_n$ for all $i$ (since otherwise $\normtwo{x-z} > \beta_n$).
  Therefore,
  \begin{align*}
      p_W(x) 
      &= s_W(x) + \lambda B(x) 
      \leq (\normtwo{w}^2+\lambda(2+\beta_n)^2)\beta_n^2
      = \frac{1}{2},\,
  \end{align*}
where the last inequality is due to our choice of $\beta_n$ in~\eqref{eq:deg4:subsetsum:beta}. This implies $f_W(x)=1$, as claimed.
\end{proof}

\begin{proofof}{\cref{prop:deg4:subsetsum:densities}}
  \cref{claim:deg4:subsetsum:farfromall,claim:deg4:subsetsum:closetobad} together imply (since $1/(4\normtwo{w} \gg \alpha_n$) that if $x\in\R^n$ is not within $\lp[2]$ distance $\alpha_n$ of a solution $z\in\{-1,1\}^n$ of the \textsf{Subset-Sum} instance $W$, then $f_W(x) = -1$. Moreover,~\cref{claim:deg4:subsetsum:closetogood} ensures that any $x\in\R^n$ which is within $\lp[2]$ distance $\beta_n$ of a solution $z\in\{-1,1\}^n$ of the \textsf{Subset-Sum} instance $W$ satisfies $f_W(x) = 1$.
  
  Therefore, as $\alpha_n = o(1)$ and any two points in $\{-1,1\}^n$ are at least at unit $\lp[2]$ distance, the satisfying assignments $f_W^{-1}(1)$ are disjoint sets $S_1,\dots, S_K$ (one around each of the $K$ satisfying assignments $z_1,\dots, z_K\in\{-1,1\}^n$ of the \textsf{Subset-Sum} instance) such that, for all $1\leq k\leq K$,
  \begin{equation*}
          B^2_n( z_k, \beta_n )\subseteq S_k \subseteq B^2_n( z_k, \alpha_n )\,.
  \end{equation*}
  This gives the first part of item (2) of the proposition.  
  
  For items (1) and (3) (and the last part of item~(2)), by symmetry it suffices to consider the case $z_k=1_n$: we want to estimate the Gaussian measure of the $\lp[2]$ ball of radius $r$, $\phi(B^2_n(1_n, r))$. For $r = O(1/\sqrt{n})$, we claim that this is within constant factors of $\phi(1_n)\cdot \operatorname*{Leb}(B^2_n(1_n,r))$, i.e.,~we claim that
  \[
      \phi(B^2_n(1_n, r )) = \bigTheta{ \operatorname*{Leb}(B^2_n(1_n,r))\cdot \frac{1}{(2\pi)^{n/2}} e^{-\normtwo{1_n}^2/2} } = 
      \bigTheta{ \frac{r^n}{2^{n/2}\Gamma(n/2+1)} e^{-n/2} }\,.
  \]
  This is because, for such a small radius $r$, the Gaussian density changes by at most a constant factor within the radius-$r$ $\lp[2]$ ball centered at $1_n$: indeed, for any $z$ in this ball, $\normtwo{z} = \sqrt{n}\pm r$, so that
  $
   e^{-\normtwo{z}^2/2} = e^{-n/2\pm O(1)} = \bigTheta{e^{-n/2}} 
  $. Therefore, by the above we have
  \begin{equation*}
          \bigOmega{\beta_n^n\frac{e^n}{\sqrt{n}\cdot n^{n/2}}} \leq \phi(S_k) \leq \bigO{\alpha_n^n\frac{e^n}{\sqrt{n}\cdot n^{n/2}}}\,.
  \end{equation*}
  From~\cref{eq:deg4:subsetsum:alpha,eq:deg4:subsetsum:beta} and our setting of $\lambda=cn\normtwo{w}^2$, we get that $\alpha_n,\beta_n$ are respectively $\frac{1}{\sqrt{8\lambda}}+\bigO{\frac{1}{\lambda}}$ and $\frac{1}{\sqrt{8\lambda}}-\bigO{\frac{1}{\lambda}}$, from which, because $\lambda \gg n^2$, we get
  \begin{equation}
          \alpha_n^n,\beta_n^n = \frac{1}{(8\lambda)^{n/2}}\left(1\pm \bigO{ \frac{1}{\sqrt{\lambda} } }\right) = \bigTheta{\frac{1}{(8\lambda)^{n/2}}}.
  \end{equation}
  This shows that all measures $\phi(S_k)$ are within a constant factor of each other. Moreover, under the Gaussian measure $\phi$ over $\R^n$, the above along with our bound $\norminf{w}\leq 2^{n^2}$ and the choice of $\lambda=
  cn\max(\normtwo{w}^2,n)$ implies that
  \begin{equation}
          \phi(S_k) = \frac{1}{2^{O(n^3)}} \,,
  \end{equation}
  as $\lambda = O(n^{2}2^{2n^2})$.
  
  This establishes item~1 of \cref{prop:deg4:subsetsum:densities}, so it remains only to show that item~3 holds. To establish this last item (i.e., that for any $k$ one can efficiently, given any $x^\ast\in S_k$, sample according to the Gaussian measure restricted to $S_k$) it suffices to observe that (i)~given such a $x^\ast$, it is immediate to find the corresponding satisfying assignment $z_k\in\{-1,1\}^n$ (by rounding the coordinates); (ii)~efficient sampling uniformly from $B^2_n( z_k, \alpha_n )$ can be done using elementary techniques; and (iii)~since the volume of $S_k$ is within a constant factor of the volume of $B^2_n( z_k, \alpha_n )$ and $f_W$ can be efficiently evaluated, and the Gaussian pdf can be easily computed on any point in $S_k$ and this value varies only by an $O(1)$ factor between any two points in $S_k$, rejection sampling allows us to sample according to the Gaussian measure restricted to $S_k$ in an efficient way.
  This concludes the proof of \cref{prop:deg4:subsetsum:densities} and hence also of \cref{thm:neg-deg4-2}.
\end{proofof}

\bibliography{allrefs}

\appendix

\section{Hardness of distribution learning with respect to arbitrary background distributions} \label{ap:intractable}
In this appendix we justify our focus on structured continuous distributions (namely Gaussian and log-concave) by proving that for general distributions ${\cal D}$,  learning $\calD_f$ even for very simple functions $f$ (indeed, linear threshold functions) is computationally hard. 

We actually prove an even stronger hardness result, by assuming both that 

\begin{itemize}

\item the background distribution $\calD$ is
provided to the learning algorithm via an evaluation oracle to a function $\tilde{\calD}$ such that on any input $x$, $ \tilde{\calD}(x)$ is proportional to $\calD(x)$; and that
\item the algorithm has access to an exact sampler for $\calD$. 

\end{itemize}

(This is at least as strong as the model we use for our positive
results on log-concave distributions, presented in~\cref{sec:preliminaries},
in which the learner only receives an evaluation oracle for some measure
which is a rescaling of the probability density function.)  We show that, even in this
setting, any distribution learning algorithm with respect to an arbitrary
background distribution $\calD$ may need to take exponential time, even
if the function $f$ is guaranteed to be a halfspace \emph{which is provided to  the learning algorithm}.

 To see
this, suppose that the distribution $\calD$ is $\calU_g$ (the uniform distribution over $\bn$ restricted to the satisfying assignments of $g$), where $g$ is a halfspace whose description is provided to the algorithm. We note that the distribution $\calD$ satisfies both our hypothesis. Namely, 
\begin{itemize}
\item The 0/1-valued function $g(x)$ is proportional to $\calD(x)$, and $g(x)$ can be efficiently computed. (In fact, the result of Dyer~\cite{Dyer:03} (and earlier work of Morris and Sinclair~\cite{MorrisSinclair:04}) also establishes that $\calD(x)$ itself can be efficiently evaluated to a multiplicative factor of $(1+\eps)$ in $\poly(n/\eps)$ time.)
\item By the result of Dyer~\cite{Dyer:03}), there is an efficient  algorithm to exactly sample from $\calD$.
\end{itemize}
Now let us define $h(x) = f(x) \wedge g(x)$, so $h$ is an intersection of two halfspaces. Note that the distribution $\calU_h$ is exactly $\calD_f$. However, under Assumption~\ref{assumption:crypto:hardness}, it was shown in~\cite{DDS:15} that there is no  algorithm with running time $2^{n^{\delta}}$ (for some fixed $\delta>0$) to learn $\calU_h$ given samples from the distribution (Theorem~1.5 of \cite{DDS:15}, see also Theorem~62 of \cite{DDS-LSA-full}). It can be verified that this hardness result continues to hold even if the description of both halfspaces $f$ and $g$ is given to the algorithm. This implies that under Assumption~\ref{assumption:crypto:hardness}, there is no  algorithm with running time $2^{n^{\delta}}$ to learn $\calD_f$ given samples from the distribution. Now, since $g$ is known to the algorithm, by the above discussion the algorithm can efficiently (i)~sample from $\calD$, and (ii)~evaluate the function $g(x)$ which is proportional to $\calD(x)$. This shows that for arbitrary  distributions $\calD$ over the discrete cube, learning $\calD_f$ is computationally hard for a halfspace $f$ even when we have access to a sampler for $\calD$ as well as an evaluation oracle to $\calD$. 

Finally, observe that this example straightforwardly generalizes 
to continuous probability distributions by replacing the points of $\{-1,1\}^n$ in the support by disjoint (suitably small) balls centered at those points, and making $\calD$ uniform on the union of these 
disjoint balls (akin to the constructions of \Cref{sec:lowerbounds}).

\section{Smooth version of $\tilde{\mu}$} \label{ap:smooth}

For technical reasons related to the details of the literature on sampling from log-concave distributions, it is helpful for us to be able to work with a ``cleaned-up'' version of the measure $\tilde{\mu}$, which we denote $\tilde{\mu}_1$.  The associated distribution $\mu_1$ is entirely supported on ${\cal B}(0,n^C)$, puts at least some minimum amount of mass on every point in ${\cal B}(0,n^C)$, and is statistically very close to the actual distribution $\mu$.  The precise statement we will use is stated and proved below.

\begin{theorem}~\label{thm:smooth}
Let $\tilde{\mu}$ be a measure such that the associated density $\mu$ satisfies conditions {\bf C1}, {\bf C2} and {\bf C3} from~\Cref{sec:preliminaries}.
Let $\tilde{\mu}_1\colon \R^n \to (0,\infty)$ be the measure which is defined as follows: 
\[
\tilde{\mu}_1(x) = \begin{cases} 0 &\textrm{ if }x\not \in B^2(0,n^C) \\ 
\tilde{\mu}(x) + 2^{-n^C-4n} \cdot n^{-Cn} &\textrm{ otherwise }\\
\end{cases}
\]
Let $\mu_1$ be the density corresponding to $\tilde{\mu}_1$. Then, 
\begin{enumerate}
\item $\mu_1$ is log-concave. 
\item For any $x \in B^2(0,n^C)$, $\mu_1(x) \ge 2^{-2n^C  -5n-3} \cdot n^{-{2}Cn}$
\item $\Vert \mu  - \mu_1 \Vert_1 \le 2^{-\frac{3n}{2}}$. 
\end{enumerate}
\end{theorem}

\begin{proof}
The first item follows simply from the fact that restricting $\tilde{\mu}$ to a convex set preserves log-concavity as does adding a constant function.

To prove the next item, observe that
\begin{align}
\int_{x} \tilde{\mu}(x)  \le& 2 \cdot \int_{x} \tilde{\mu}_T(x) = 2\cdot \int_{x \in B^2(0,n^C)} \tilde{\mu}_T(x) + 2 \int_{x \not \in B^2(0,n^C)} \tilde{\mu}_T(x) \ \ \textrm{(uses \textbf{C3})} \nonumber\\
\leq& 2\cdot \int_{x \in B^2(0,n^C)} \tilde{\mu}_T(x) + 2  \int_{x \not \in B^2(0,n^C)} \tilde{\mu}(x)  \tag{since $\tilde{\mu}(x) \leq \mu(x)$ for all $x$}\nonumber \\
\leq& 2\cdot \int_{x \in B^2(0,n^C)} \tilde{\mu}_T(x) + 2^{1-Cn} \cdot \int_{x} \tilde{\mu}(x)  \ \ \textrm{(uses \textbf{C1})}  \label{eq:lb2}
\end{align}
We now make two observations: 
\begin{enumerate}
\item[(i)] By {\bf C2}, we have that $\tilde{\mu}_T(x)$ is at most $2^{n^C}$ for all $x$. Since the volume of $B^2(0,n^C)$ is at most $2^{n} \cdot n^{Cn}$, the first integral in \Cref{eq:lb2} is at most 
$2^{n + n^C} \cdot n^{Cn}$. 
\item Since $n \ge 1$ and $C \ge 2$, $2^{1-Cn} \le 1/2$. 
\end{enumerate}

Combining these two observations with (\ref{eq:lb2}),
we have 
$
\int_x \tilde{\mu}(x) \le 2^{n^C + n +2} \cdot n^{Cn},
$ 
and hence $\int_x \tilde{\mu}_1(x) \leq 2^{n^C + n + 3}$ (with room to spare).
This immediately implies that for any $x \in B^2(0,n^C)$, 
\[
\mu_1(x) = \frac{\tilde{\mu}_1(x)}{\int_x \tilde{\mu}_1(x)} \ge  2^{-2n^C -5n -3} \cdot n^{-2Cn}. 
\]
This proves the second item. To prove the third item, first observe that if $\mu_T$ denotes the density induces by the measure $\tilde{\mu}_T$, then 
\[
\max_{x} \mu_T(x) = \max_x \frac{\tilde{\mu}_T(x)}{\int_{x} \tilde{\mu}_T(x)} \le \max_x \frac{2\tilde{\mu}(x)}{\int_{x} \tilde{\mu}(x)} \le 2^{n+1}. 
\]

Here the penultimate inequality uses that $\tilde{\mu}_T(x) \le \tilde{\mu}(x)$ as well as condition \textbf{C3}, and the last inequality uses condition {\bf C2}. Consequently, if $\mathcal{A} = \{x: \mu_T(x) \not =0\}$, then since $\mu_T$ is a distribution, it must hold that $\mathsf{Vol}(\mathcal{A}) \ge 2^{-n-1}$.  We thus have that
\[
\int_x \tilde{\mu}_T(x) \ge \int_{x: \mu_T(x) \not =0} \tilde{\mu}_T(x) \ge \min_{x: \mu_T(x) \not =0} \tilde{\mu}_T(x) \cdot \mathsf{Vol}(\mathcal{A}) \ge 2^{-n-1-n^C},
\]
where the last inequality is by the definition of $\tilde{\mu}_T$ as given in condition {\bf C3} (truncating $n^C$ bits after the binary point).
Again using that $\tilde{\mu}(x) \ge \tilde{\mu}_T(x)$, we obtain 
\begin{equation}~\label{eq:integral-lb}
\int_x \tilde{\mu}(x) \ge 2^{-n-1-n^C}. 
\end{equation}
Further, we also have that 
\begin{eqnarray}
\left|\int_x\tilde{\mu}(x) - \tilde{\mu}_1(x) \right | &\le& \int_{x \not \in B^2(0,n^C)} |\tilde{\mu}(x)| +  \int_{x  \in B^2(0,n^C)} |2^{-n^C -4n} \cdot n^{-Cn}| \nonumber  \\
&\le& \int_{x} \tilde{\mu}(x) \cdot 2^{-Cn} + 2^{-n^C -3n}  \ \textrm{(uses \textbf{C1})} \nonumber \\
&\le& \int_{x} \tilde{\mu}(x) (2^{-Cn} + 2^{-2n +1}), \label{eq:integral-lb-2}
\end{eqnarray}
where the last inequality is by (\ref{eq:integral-lb}). With this, we are ready to prove the third item. 

\begin{eqnarray}
\int_{x} |\mu(x) - \mu_1(x)| &=& \int_{x \in B^2(0,n^C)} |\mu(x) - \mu_1(x)| + \int_{x \not \in B^2(0,n^C)} |\mu(x) - \mu_1(x)| \nonumber 
\\ 
&=& \int_{x \in B^2(0,n^C)} |\mu(x) - \mu_1(x)| + \int_{x \not \in B^2(0,n^C)} |\mu(x) | \  \ \textrm{(by definition of }\mu_1\textrm{)} \nonumber \\
&\le& \int_{x \in B^2(0,n^C)} |\mu(x) - \mu_1(x)| + 2^{-Cn}  \  \ \textrm{(uses \textbf{C1})} \nonumber \\&\le& \int_{x \in B^2(0,n^C)} \bigg|\frac{\tilde{\mu}(x)}{\int_x \tilde{\mu}(x)} - \frac{\tilde{\mu}_1(x)}{\int_x \tilde{\mu}(x)}\bigg| +\int_{x \in B^2(0,n^C)} \bigg|\frac{\tilde{\mu}_1(x)}{\int_x \tilde{\mu}_1(x)} - \frac{\tilde{\mu}_1(x)}{\int_x \tilde{\mu}(x)}\bigg| + 2^{-Cn}. \label{eq:integral-lb-3}
\end{eqnarray}
Here the last inequality uses the definitions of $\mu$ and $\mu_1$ and the triangle inequality. We now bound the terms in \Cref{eq:integral-lb-3} individually.  We start with the first term. 

\begin{eqnarray}
\int_{x \in B^2(0,n^C)} \bigg|\frac{\tilde{\mu}(x)}{\int_x \tilde{\mu}(x)} - \frac{\tilde{\mu}_1(x)}{\int_x \tilde{\mu}(x)}\bigg| &\le \frac{2^{-n^C -4n} \cdot n^{-Cn}}{2^{-n-1-n^C}} \cdot \mathsf{Vol}(B(0,n^C))  \ \textrm{ (uses \ref{eq:integral-lb})} \nonumber\\
&\le \frac{2^{-n^C -4n} \cdot n^{-Cn}}{2^{-n-1-n^C}} \cdot 2^n \cdot n^{Cn} \le 2^{-2n}.  \label{eq:integral-lb-4}
\end{eqnarray}
To bound the second term,  we have
\[
\int_{x \in B^2(0,n^C)} \bigg| \frac{\tilde{\mu}_1(x)}{\int_x \tilde{\mu}_1(x)} - \frac{\tilde{\mu}_1(x)}{\int_x \tilde{\mu}(x)}\bigg| \le \int_{x\in B^2(0,n^C)} \frac{\tilde{\mu}_1(x)}{\int_x \tilde{\mu}_1(x)} \cdot (2^{-Cn} + 2^{-2n+1}) \le 2^{-Cn} + 2^{-2n+1}. 
\]
Combining this with (\ref{eq:integral-lb-4}) and plugging back into (\ref{eq:integral-lb-3}), we get 
\[
\int_{x} |\mu(x) - \mu_1(x)| \le 2^{-Cn} + 2^{-2n} + 2^{-Cn} + 2^{-2n+1} \le 2^{-\frac{3n}{2}}. 
\]
The above assumes that $n \ge 5$. This finishes the proof.  

\end{proof}

\end{document}